\pdfoutput=1
\documentclass[11pt, twoside, table]{predoc2}
\usepackage{microtype}
\usepackage{graphicx}
\usepackage[export]{adjustbox}
\usepackage{tikz}
\usepackage{fancyvrb,newverbs,xcolor} 
\usepackage[shortlabels,inline]{enumitem}
\usepackage{longtable}
\usepackage{booktabs}
\usepackage{multirow}
\usepackage{pdflscape}
\usepackage[title]{appendix}
\usepackage{float}
\newcommand\mybox[2][]{\tikz[overlay]\node[fill=gray!50,inner sep=2pt, anchor=text, rectangle, rounded corners=1mm,#1] {#2};\phantom{#2}}
\usetikzlibrary{shapes,shapes.arrows,arrows,arrows.meta,fit,positioning,tikzmark,automata, graphs, graphs.standard}
\tikzset{
    state/.style={ellipse, draw, minimum width = 0.7 cm},
    point/.style={circle, draw, inner sep=0.04cm,fill,node contents={}},
    directed/.style={-Latex,auto,node distance =1 cm and 1 cm,semithick},
    bidirected/.style={Latex-Latex,dashed},
    el/.style = {inner sep=2pt, align=left, sloped}
}

\title{Supporting Bayesian modelling workflows with iterative filtering for multiverse analysis}
\author[1]{Anna Elisabeth Riha}
\author[1]{Nikolas Siccha}
\author[2]{Antti Oulasvirta}
\author[1]{Aki Vehtari}
\affil[1]{Department of Computer Science, Aalto University, Finland}
\affil[2]{Department of Information and Communications Engineering, Aalto University, Finland}

\makeatletter
\def\runauthor{Riha et al.}
\def\runtitle{Supporting Bayesian modelling workflows}
\makeatother

\begin{document}
\maketitle

\begin{abstract}
When building statistical models for Bayesian data analysis tasks, required and optional iterative adjustments and different modelling choices can give rise to numerous candidate models. 
In particular, checks and evaluations throughout the modelling process can motivate changes to an existing model or the consideration of alternative models to ultimately obtain models of sufficient quality for the problem at hand. 
Additionally, failing to consider alternative models can lead to overconfidence in the predictive or inferential ability of a chosen model. 
The search for suitable models requires modellers to work with multiple models without jeopardising the validity of their results.
Multiverse analysis offers a framework for transparent creation of multiple models at once based on different sensible modelling choices, but the number of candidate models arising in the combination of iterations and possible modelling choices can become overwhelming in practice.
Motivated by these challenges, this work proposes iterative filtering for multiverse analysis to support efficient and consistent assessment of multiple models and meaningful filtering towards fewer models of higher quality across different modelling contexts. 
Given that causal constraints have been taken into account, we show how multiverse analysis can be combined with recommendations from established Bayesian modelling workflows to identify promising candidate models by assessing predictive abilities and, if needed, tending to computational issues. 
We illustrate our suggested approach in different realistic modelling scenarios using real data examples.
\end{abstract}


\thispagestyle{empty}

\section{Introduction}

When specifying statistical models to analyse data, a modeller faces a range of tasks, some of which can be ambiguous or are tackled ad-hoc in practice. 
In the Bayesian framework, the desire for principled pathways from initial models to reliable inferences and decision-making motivated the notion of Bayesian workflows \citep[see, e.g.,][]{savage_what_2016, gabry_visualization_2019, betancourt_towards_2020, gelman_bayesian_2020, martin_bayesian_2021}. 
These are explicit collections of required and recommended steps to ensure valid model development. 
The guidelines provided by Bayesian workflows can request the modeller to navigate potentially many iterations within and across models.

Consider the task of applying Bayesian models to evaluate the effect of a treatment in a randomised controlled trial. 
This requires the modeller to choose covariates, hierarchical model components, and distributional families for the observations and (marginal) priors.
The modeller has to iterate, for example, if a model is of interest but not usable because computational issues result in posterior samples of insufficient quality. 
To inform decisions and understand the effect of treatment in the given study, the modeller investigates the posterior results for the quantity of interest as well as the predictive abilities of the models.
Even models that are substantively and practically relevant may turn out to be inferior candidates, for example, due to poor predictive performance.
While using one model without analysing alternatives risks an over- or underestimated effect of the treatment, investigating numerous models can result in an overwhelming amount of additional joint evaluation and model comparisons. 
Many candidate models are possible, but not all of these will be substantively relevant for assessing the effect of the treatment.
As we demonstrate in Section \ref{subsec:epilepsy-ppc} and Section \ref{subsec:epilepsy-comp}, different models can also imply different conclusions about the treatment under evaluation.
Various iterative adjustments in conjunction with different possible modelling choices unavoidably require the modeller to examine multiple models. 
However, working efficiently and consistently with sets of models is not straightforward; especially when iteration alters the set of considered models. 

Motivated by these challenges, we explore how to support modellers throughout iterative Bayesian modelling by combining workflow tasks with multiverse analysis: the joint and transparent creation of multiple models based on different sensible modelling choices \citep[see, e.g.,][]{steegen_increasing_2016, simonsohn_specification_2020}. 
\begin{figure}[t]
    \centering
    \begin{tikzpicture}
  \node[draw, rounded rectangle, minimum width=50mm,thick] at (0.6,1.4) {\footnotesize{All models}};
  \path [draw=gray,thick] (-1.3,1.15) edge (-1.3,0.6);
  \path [draw=gray,thick] (-1,1.15) edge (-1,0.6);
  \path [draw=gray,thick] (-0.7,1.15) edge (-0.7,0.6);
  \path [draw=red,dashed,thick] (-0.4,1.15) edge (-0.4,0.6);
  \path [draw=red,dashed,thick] (-0.1,1.15) edge (-0.1,0.6);
  \path [draw=red,dashed,thick] (0.2,1.15) edge (0.2,0.6);
  \path [draw=red,dashed,thick] (0.5,1.15) edge (0.5,0.6);
  \path [draw=red,dashed,thick] (0.8,1.15) edge (0.8,0.6);
  \path [draw=red,dashed,thick] (1.1,1.15) edge (1.1,0.6);
  \path [draw=gray,thick] (1.4,1.15) edge (1.4,0.6);
  \path [draw=gray,thick] (1.7,1.15) edge (1.7,0.6);
  \path [draw=red,dashed,thick] (2,1.15) edge (2,0.6);
  \path [draw=red,dashed,thick] (2.3,1.15) edge (2.3,0.6);
  \path [draw=red,dashed,thick] (2.6,1.15) edge (2.6,0.6);
  \node[draw, rounded rectangle, minimum width=50mm, thick] at (1.2,0.3) {\footnotesize{Filtering by predictive abilities}
  
  \scriptsize{(Section \ref{subsec:pred-eval-filtering})}};
  \path [draw=gray,thick] (-1.3,0) edge (-1.3,-2);
  \path [draw=gray,thick] (-1,0) edge (-1,-2);
  \path [draw=gray,thick] (-0.7,0) edge (-0.7,-2);
  \path [draw=red,dashed,thick] (-0.4,0) edge (-0.4,-0.5);
  \path [draw=red,dashed,thick] (-0.1,0) edge (-0.1,-0.35);
  \path [draw=red,dashed,thick] (0.2,0) edge (0.2,-0.35);
  \path [draw=red,dashed,thick] (0.5,0) edge (0.5,-0.35);
  \path [draw=red,dashed,thick] (0.8,0) edge (0.8,-0.35);
  \path [draw=red,dashed,thick] (1.1,0) edge (1.1,-0.35);
  \node [rounded rectangle, minimum height=13mm, minimum width=37mm, draw, thick] at (1.1,-1) {};
  \node[draw] at (0.16,-1) (here1) {\footnotesize Check};
  \node[draw] at (2,-1) (here2) {\footnotesize Change};
  \path [directed] (here1) edge[bend left=50]  (here2);
  \path [directed] (here2) edge[bend right=-50]  (here1);
  % 9 for preprint 
  \node[rectangle, text width=34mm, minimum height=20mm, draw, thick, align=center, rounded corners] at (-3.6,-1) {\footnotesize{Computation checks} \scriptsize{
            \begin{enumerate}[nosep]
            \item[\textbullet] convergence 
            \item[\textbullet] efficiency
            \item[\textbullet] reliability
            \item[\textbullet] $\cdots$
        \end{enumerate}}
        
        (Section \ref{subsec:computation-inference})};
  \node[rectangle, text width=34mm, minimum height=20mm, draw, thick, align=center, rounded corners] at (6,-1) {\footnotesize{Iterative improvement} \scriptsize{
        \begin{enumerate}[nosep]
            \item[\textbullet] modify model 
            \item[\textbullet] modify computation
            \item[\textbullet] extend models
            \item[\textbullet] $\cdots$
        \end{enumerate}}
        
        (Section \ref{subsec:iterating})};
  % loop
  \path [directed] (here1) edge[bend left=50]  (here2);
  \path [directed] (here2) edge[bend right=-50]  (here1);
  % dotted paths 
  \path [draw=black, dotted, thick] (2.6,-0.8) to (4.2,-0.05);
  \path [draw=black, dotted, thick] (2.6,-1.2) to (4.2,-1.9);
  \path [draw=black, dotted, thick] (-0.4,-0.8) to (-1.8,-0.05);
  \path [draw=black, dotted, thick] (-0.4,-1.2) to (-1.8,-1.9);
  \node[draw, rounded rectangle,thick] at (1.2,-2.3) {\footnotesize{Filtering by predictive abilities} \scriptsize{(Section \ref{subsec:pred-eval-filtering})}};
  \path [draw=gray,thick] (-0.4,-1.52) edge (-0.4,-2);
  \path [draw=gray,thick] (-0.1,-1.64) edge (-0.1,-2);
  \path [draw=gray,thick] (0.2,-1.64) edge (0.2,-2);
  % lowest paths
  \path [directed,draw=gray,thick] (-1.3,-2.6) to (-1.3,-3.25);
  \path [directed,draw=gray,thick] (-1,-2.6) to (-1,-3.25);
  \path [directed,draw=gray,thick] (-0.7,-2.6) to (-0.7,-3.25);
  \path [directed,draw=gray,thick] (-0.4,-2.6) to (-0.4,-3.25);
  \node[draw, rounded rectangle, thick] at (-0.2,-3.5) {\footnotesize{Filtered set of models}};
\end{tikzpicture}
    \vspace*{-1.5\baselineskip}
    \caption{Workflow for iterative filtering with multiple models. We suggest to use (1) predictive abilities, as well as (2) computation and model checking and iterative improvements, to filter many models towards fewer models of higher quality. Each vertical line represents one model. Models indicated by dashed \textcolor{red}{red} lines are problematic, for example, due to computational issues. Such models are either dropped after assessing predictive abilities or require iterative adjustments which can make them valid candidates.}
    \label{fig:approach}
\end{figure}
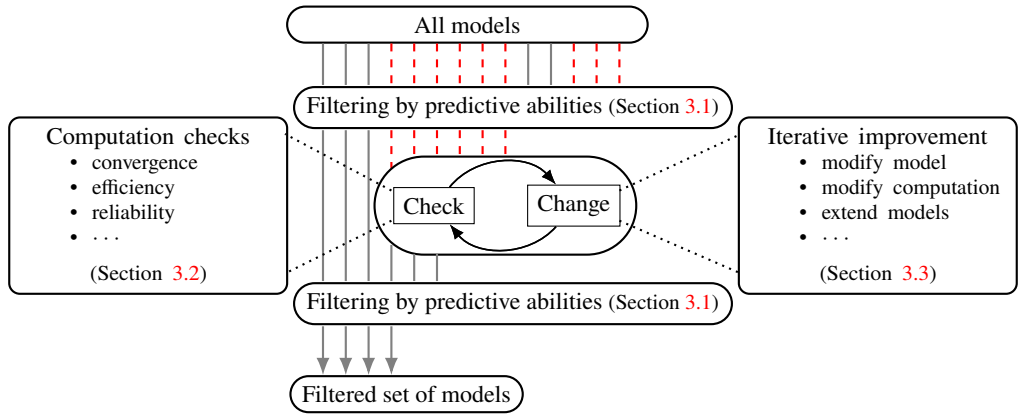
In particular, we 
\begin{enumerate}[nosep]
    \item[\textbullet] propose iterative filtering for multiverse analysis as a tool for Bayesian modelling workflows that reduces large sets of candidate models using assessment of predictive abilities and evaluation of computation (see Figure \ref{fig:approach} and Section \ref{sec:multiverse-workflow})
    \item[\textbullet] investigate desirable properties of iterative Bayesian modelling workflows that can serve as a practical 
    foundation for developing tools to support modellers (see Section \ref{sec:support-iter})
    \item[\textbullet] illustrate the suggested filtering approach and the advantages of jointly evaluating models and computation in real data examples (see Section \ref{sec:case-studies})  
\end{enumerate}

\subsection{Bayesian workflows and iterative modelling}\label{subsec:intro-workflow}
A Bayesian modelling workflow is a systematic collection of iterative steps and sequential decisions for building, evaluating, and comparing models that provides principled pathways towards valid models within a Bayesian framework \citep[see, e.g.,][]{savage_what_2016, gabry_visualization_2019, betancourt_towards_2020, gelman_bayesian_2020, martin_bayesian_2021}.
Several aspects of Bayesian modelling necessitate iteration within and across models. 
For example, model checking can reveal a misspecified model or inherent aspects of the modelling task, like the complexity of a model, can complicate numerical approximations. 
Modifications can be essential for reliable posterior inferences, or to obtain posterior samples in the first place. 
Iteration can be desirable, for instance, if predictive performance is not sufficient or new data becomes available.
The iterative facet of model building also implies that more resources like time or modelling ideas can motivate ``remodelling'' \citep[][p. 418]{bernardo_bayesian_1994}, that is, iteration towards improved models.

Building models iteratively has been a long-standing topic of concern and interest, especially in the Bayesian paradigm, potentially inspired by its ability to incorporate new evidence and its reliance on numerical approximations \citep[see, e.g.,][]{box_science_1976,  leamer_specification_1978,tukey_we_1980, bernardo_bayesian_1994,gelman_bayesian_2020}. 
\cite{box_science_1976} identifies learning as the main motivation of any modelling endeavour and, consequently, describes it as a process that unfolds in a ``motivated iteration between theory and practice'' \citep[][p. 791]{box_science_1976}. 
Adopting the hypothetico-deductive perspective discussed by \citet{gelman_philosophy_2012}, models encapsulate hypotheses from which implications for observed data can be deduced, for example, by obtaining posterior replicates for the outcome of interest. 
These hypotheses are to be questioned, if, for example, the model is not providing sufficiently accurate predictions or is failing to capture essential aspects of the phenomenon of interest.
As such, modelling necessarily evolves in an iteration between formulating hypotheses and testing their implications.

Even if existing scientific knowledge suggests an ideal model, it can still be infeasible to evaluate and debug this model directly.
As discussed by \citet{gelman_bayesian_2020}, building models iteratively with increasing complexity is common in practice. 
At the same time, it can be challenging to make all considered models explicit. 
Moreover, due to the selective nature of any step-wise procedure, we inevitably risk overlooking a model that would contradict previous conclusions. 
Adjusting and refining models step-by-step seems intuitive, but it is not inherently efficient or transparent, and does not come with automatic guarantees of sufficient exploration.

\subsection{Models everywhere, all at once}\label{subsec:intro-multiverse}
Instead of hand-picking models or iteratively refining one model, multiverse analysis provides a framework for joint and transparent consideration of several candidate models \citep{steegen_increasing_2016}\footnote{Similar and related approaches have been suggested for working with sets of possible models, among them, ``factorial model comparison'' \citep[]{van_den_berg_factorial_2014}{}, ``vibration of effects'' \citep[]{patel_assessment_2015}, ``model influence analysis'' \citep[]{young_model_2017}, or ``specification curve analysis'' \citep[]{simonsohn_specification_2020}.}.
An explicit multiverse of possible results is motivated by the ``garden of forking paths'' \citep[][]{gelman_garden_2013}{}{}; the issue of arriving at different models given variations in observed data.
A ``multiverse of models'' emerges from a range of possible modelling choices \citep[][]{steegen_increasing_2016}. 

The \texttt{R}-package \texttt{multiverse} \citep{sarma_multiverse_2021} enables straightforward multiverse analyses from existing \texttt{R}-code by adding branches at decision points in a given data analysis with \texttt{multiverse::branch()}.
While simultaneously considering multiple models increases transparency and reduces the risk of overconfidence in a single model, models in a multiverse can appear equally plausible without additional assessment of the validity and relevance of models \citep[see, e.g.,][]{ simonsohn_specification_2020,hall_survey_2022}.
Challenges in investigating results of multiverse analyses motivated ideas for explorable multiverse analysis reports \citep[]{dragicevic_increasing_2019} and an interactive workflow for multiverse analysis in \citep[][]{liu_boba_2021}.
The accompanying language-agnostic tool \texttt{Boba} \citep[][]{liu_boba_2021} can distinguish models, for example, based on normalised root mean squared error.
To assist a flexible consideration of multiple Bayesian models, \citet{bernstein_multi-model_2022, bernstein_abstractions_2023} introduces multi-model probabilistic programming for the probabilistic programming language \texttt{Stan} \citep[][]{stan_development_team_stan_2023}, and explores how to navigate a multiverse of models along paths of increasing predictive performance \citep[see Figure 3.14 in][p. 141]{bernstein_abstractions_2023}.

Albeit increasing transparency, multiverses ask the modeller to keep track of many variations of models, which might itself become unmanageable \citep[see, e.g.,][]{steegen_increasing_2016, bernstein_abstractions_2023}. 
A joint consideration of numerous candidate models can overwhelm and mislead due to the increased amount of parallel model evaluations and visual biases when reporting results \citep[see][Section 7]{hall_survey_2022}.
To make use of the advantages of multiverse analysis within Bayesian workflows, the question is how to identify models of interest in potentially overwhelming and dynamic multiverses. 
In particular, modellers need tools for efficient and consistent joint evaluation of different aspects of models, including but also going beyond predictive performance.

\subsection{Our approach}
\begin{figure}[t]
    \centering
    \includegraphics[width=0.8\textwidth]{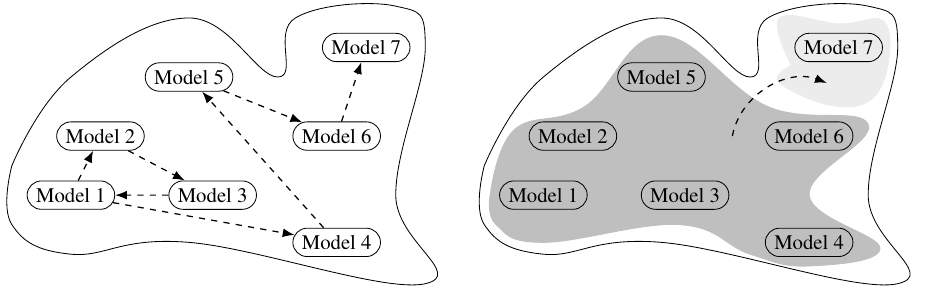}
    \caption{Exploration of model space with iterative modelling and multiverse analysis. An iterative model building procedure is illustrated on the left by models connected with dashed lines. After evaluating Model 2 and 3, we decide to move back to Model 1 and from there continue to Model 4. The \mybox[fill=gray!50]{grey} area on the right shows a set of six models created at once via multiverse analysis. Model 7 in the \mybox[fill=lightgray!20]{lighter grey} area is another model achieved, for example, by extending the initial multiverse.}
    \label{fig:multiverse-modelspace}
\end{figure}

Driven by the need to jointly investigate several models in Bayesian modelling workflows, we propose iterative filtering for multiverse analysis in connection to the three steps of a Bayesian data analysis task proposed by \citet{gelman_etal_bda3_2013}, that is, 
\begin{enumerate*}[label=(\Roman*)]
    \item model building, 
    \item conditioning on data and deriving posterior inference, and
    \item model evaluation.
\end{enumerate*}
Instead of building and evaluating models one by one in step (I), we immediately create multiple initial models using multiverse analysis, as illustrated in Figure \ref{fig:multiverse-modelspace}.
In contrast to focusing on one model until it fails, we connect step (II) and step (III) by allowing for iterative improvements of computation and models while checking and evaluating multiple models jointly. 

Specifically, we perform iterative filtering by assessing computational issues in inference, performing posterior predictive checks, evaluating predictive performance, and checking the reliability of estimates for predictive performance. 
Given that causal constraints have been taken into account, the suggested approach can reduce a set of candidates to a subset of models of higher quality, as illustrated in Figure \ref{fig:approach}.  
Iterative filtering for multiverse analysis considers a multiverse of models by design. 
Instead of building and choosing models based on opaque assessments, the suggested approach ensures transparency and provides clear evaluation criteria to systematically reduce a set of models. 
This avoids an arbitrary selection of models and allows comparisons between different modelling contexts.  

This paper is structured as follows. 
Relations between iterative modelling and multiverse analysis are discussed in Section \ref{subsec:multiverse-bayesian-models}. 
We motivate the need for meaningful filtering in Section \ref{subsec:justifyfiltering} and introduce the chosen filtering criteria in Section \ref{subsec:filtering-min-viable}. 
Iterative filtering for multiverse analysis is described in Section \ref{sec:multiverse-workflow} and connects the tasks of checking computation (see step (II) above and Section \ref{subsec:computation-inference}) as well as evaluating obtained models and their implied conclusions (see step (III) above and Section \ref{subsec:pred-eval-filtering}).
Section \ref{subsec:iterating} discusses how the suggested approach adds the possibility of iterating towards improved sets of models to the steps of a Bayesian data analysis task by \citet{gelman_etal_bda3_2013}. 
In Section \ref{sec:case-studies}, real-data examples illustrate that iterative filtering allows clear communication about modelling choices and transparent reduction towards fewer models of higher quality.
Section \ref{sec:discussion} summarises results, and discusses limitations of the presented approach as well as future directions. 

\section{Filtering a Bayesian multiverse towards minimum viable models}\label{sec:support-iter}

\subsection{Building multiple models transparently with multiverse analysis}\label{subsec:multiverse-bayesian-models}

For a given data analysis task that involves model building, potentially infinitely many candidate models are contained in a space of models $\mathcal{M}$. 
As introduced already briefly in Section \ref{subsec:intro-multiverse}, a multiverse can be defined as a finite set of models $M = \{M_1, \cdots, M_K\} \subset \mathcal{M}$, where each model $M_k$ is characterised by a unique combination of modelling choices \citep[see, e.g.,][]{steegen_increasing_2016, dragicevic_increasing_2019, bell_modeling_2022}. 
In a Bayesian framework, such choices need to be made, for example, when considering which distributional families for the observations are suitable or what different prior information can be encoded into the model.
While \cite{steegen_increasing_2016} focus on multiverses emerging due to different data preprocessing, this work revolves around multiverses of Bayesian models. 
Each model $M_k = p_k(y, \theta_k, {\rm{X}}_k)$ is defined by different selected covariates in ${\rm{X}}_k$, as well as distributions for the response $y$, $p_k(y \mid \theta_k)$, with parameters $\theta_k$, and prior distributions $p_k(\theta_k)$ for the same observed data $y$. 

Note that we do not suggest that any individual combination of modelling choices is the truth. 
Instead, we want to investigate how the breadth of analysis conclusions can be impacted by selecting only one of these combinations.
Indeed, the ability to consider them all in parallel using multiverse analysis can provide a more representative view of possible conclusions than any single model could offer.
Multiverse analysis offers a convenient framework for embedding transparency in Bayesian modelling workflows. 
It ensures accessibility of all modelling attempts and the explicit formulation of all considered modelling choices.
Moreover, the possibility of joint exploration of several models is attractive since the modeller can iterate over multiple models as opposed to trying different models sequentially.

\subsection{Filtering is needed and justified}\label{subsec:justifyfiltering}

From the perspective of Bayesian model selection and Bayesian model averaging, multiverse analysis is likely interpreted as assuming the same weight for all models a priori; a best guess by the modeller which aspects could be relevant for modelling the problem at hand. 
Meaningful filtering criteria can enable transparent evaluation and ranking of models.
This allows the modeller to update which models they consider to be more promising candidates.  

As discussed in Section \ref{subsec:intro-multiverse}, models in a multiverse can seem equally plausible and a multiverse does not directly include validations and rankings of models \citep[see, e.g.,][]{steegen_increasing_2016, dragicevic_increasing_2019, simonsohn_specification_2020}{}{}.
Acknowledging the inherent limitations of every model, it is essential to identify which models are ``importantly wrong'' \citep[][p. 792]{box_science_1976}, that is, which models are not useful in the given modelling context and decision problem. 
We can validate models in a multiverse by checking model quality and evaluate them jointly with model comparison techniques.
Filtering can reduce complexity by excluding candidates from the modelling process that turn out to be largely inferior candidates. 
Subsequently, the non-arbitrarily filtered set of models enables a more detailed investigation of implied conclusions of the remaining models.

\cite{steegen_increasing_2016}, \cite{dragicevic_increasing_2019} and \cite{hall_survey_2022} emphasise that multiverse analyses should only consider models based on ``reasonable'' or ``defensible'' modelling choices to ensure meaningful comparisons between models. 
Naturally, the definition of a reasonable modelling choice can change over time and differ between modellers and fields; see, for example, ``many-analysts'' studies presented and discussed by \cite{silberzahn_many_2018} and \cite{kummerfeld_one_2023}.
We argue that it is less ambiguous to agree on a set of filtering criteria reflecting desired properties of models than to fix a generalisable notion of justified modelling choices.
For consistent assessment of models within and across different modelling endeavours, filtering can incorporate shared indicators of model quality informed by desired model properties as well as widely accepted recommendations.

\subsection{Focusing on minimum viable Bayesian models}\label{subsec:filtering-min-viable}
Filtering a set of models in a transparent and principled way requires explicit filtering criteria. 
These criteria can be used to encode the concept of a sufficiently good model in the filtering procedure.
To define meaningful criteria, we make use of recommendations in Bayesian workflows summarised by \cite{gelman_bayesian_2020} as well as ideas of ``utility dimensions'' of Bayesian models from \cite{burkner_models_2023}. 
In particular, our approach focuses on evaluating computation and predictive abilities as minimum requirements of a viable Bayesian model because
\begin{enumerate}[nosep]
    \item[\textbullet] predictive performance is a well-established and widely adopted model selection criterion related to usefulness and practical applicability of a model \citep[see, e.g,][]{ohagan_bayesian_2004, vehtari_survey_2012, piironen_comparison_2017}{}{};
    \item[\textbullet] predictive ability is connected to parameter recoverability for causally consistent models \citep[][]{scholz_prediction_2023}{}{};
    \item[\textbullet] and evaluating convergence, efficiency, and reliability of computations as well as obtained estimates is necessary to ensure the validity of our results when relying on approximations and resampling methods \citep[][]{gelman_bayesian_2020, burkner_models_2023}. 
\end{enumerate}
Multiverse analysis enables model building with several models, as opposed to considering one model at a time. 
When combined with the above filtering criteria, we can make use of the increased transparency while efficiently focusing required iterative adjustments on smaller sets of models.  
\citet{bernstein_abstractions_2023} searches a multiverse along paths of increasing predictive performance in a network of connected models. 
Similarly, predictive abilities are an important characteristic of a model in our suggested approach. 
Iterative filtering involves the evaluation of predictive performance in tandem with other criteria of model quality, as well as the implied conclusions about the quantities of interest. 
The suggested approach conserves some advantages of multiverse analysis by providing joint and transparent assessment of different aspects of multiple models. 
To work towards fewer models of higher quality, iterative filtering is not primarily concerned with a path through a network of models but instead with the distinction between inferior and more promising subsets of models therein.

\section{Integrating iterative filtering into Bayesian workflows}\label{sec:multiverse-workflow}
As opposed to repeatedly considering one model at a time until it fails, we suggest a different approach by simultaneously evaluating multiple possible models, as illustrated in Figure \ref{fig:approach}.

\subsection{Filtering by predictive abilities}\label{subsec:pred-eval-filtering}
For a set of models transparently created with multiverse analysis, the modeller obtains posterior results for the different models by conditioning on the observed data. 
In line with the goal of identifying useful models for applications and decision problems, a filtering effort starts and concludes by comparing predictive abilities of the models (see ``Filtering by predictive abilities'' in Figure \ref{fig:approach}). 
Tangentially to model selection, the modeller takes the negative view: having produced a pool of candidate models, the goal is to remove all models \textit{not} satisfying a given set of criteria reflecting predictive abilities of models. 
In particular, the proposed approach employs visual posterior predictive checks \citep[PPC;][]{gelman_etal_bda3_2013, gabry_visualization_2019}, predictive performance evaluated by expected log point-wise predictive densities \citep[\textrm{elpd},][]{vehtari_practical_2017}{}{} and the reliability of estimates of \textrm{elpd} \citep{sivula_uncertainty_2022}.

\subsubsection*{Investigating misspecification with posterior predictive checks}

To identify useful models, we need to investigate the hypotheses encapsulated in each model and, in particular, their implications for the observations. 
This process is often summarised as model checking and comprises an integral part of Bayesian data analysis \citep[see, e.g.,][]{gelman_philosophy_2012, gelman_bayesian_2020}{}. 
\citet{gabry_visualization_2019} explore and summarise the effective use of different visualisations for model checking and other essential parts of Bayesian modelling workflows.
\citet{sailynoja_graphical_2022} show how confidence bands for the empirical cumulative distribution function (ECDF) of probability integral transformation values can enhance posterior predictive checking.  
For example, visual posterior predictive checks with ECDFs can be used to identify misspecified models by comparing observed data and posterior replicates of a given model (see examples in Figure \ref{fig:epi-1-initial-ecdf-model22-21} in Section \ref{subsec:epilepsy-ppc} and Figure \ref{fig:epi-1-initial-ppc-ecdf-model-17-24} in Appendix \ref{app:casestudy-epi-1-ppc}).

\subsubsection*{Assessing predictive performance with expected log pointwise predictive density}

To assess and compare predictive performance for a set of models, we evaluate the predictive density of each model using the log score as a default utility \citep[]{geisser_predictive_1979,bernardo_bayesian_1994}. 
Following \cite{geisser_predictive_1979}, \cite{vehtari_practical_2017} denote the expected log point-wise predictive density (\textrm{elpd}) for one model and $n$ new observations as
\begin{align}\label{eq:pointwise-pred-density}
    \textrm{elpd} = \sum_{i=1}^n \int p_t(\tilde{y_i}) \log p(\tilde{y_i} \mid y) d \tilde{y_i},
\end{align}
where $p_t(\tilde{y_i})$ is the unknown true data generating process for the new $\tilde{y_i}$.
Since we do not have access to $p_t(\tilde{y_i})$, \textrm{elpd} can be estimated with cross-validation \citep[]{geisser_predictive_1975,geisser_predictive_1979}, for example, using importance sampling following \cite{gelfand_model_1992} and \cite{gelfand_model_1995} to compute approximate leave-one-out cross-validation (LOO-CV).
In particular, we obtain estimates $\widehat{\mathrm{elpd}}$ with LOO-CV and Pareto smoothed importance sampling \citep[PSIS,][]{vehtari_pareto_2024} to stabilise the importance weights for each data point \citep[][]{vehtari_practical_2017}. 

To compare models in a set of models $M = \{M_1, \cdots, M_K\}$, we make use of the difference in estimated $\mathrm{elpd}^k$ of each model $M_k$ compared to the model with the highest estimated \textrm{elpd}.
Based on estimates $\widehat{\mathrm{elpd}}$ for all models in $M$, the difference $\Delta \widehat{\mathrm{elpd}}^k$ of model $M_k$ is obtained as
\begin{align}\label{eq:elpd-difference}
    \Delta \widehat{\mathrm{elpd}}^k = \widehat{\mathrm{elpd}}^k - \max\left(\widehat{\mathrm{elpd}}\right).    
\end{align}
For the model $M_j$ in $M = \{M_1, \cdots, M_K\}$ with highest $\widehat{\mathrm{elpd}}$, that is, the model with the best predictive performance according to estimated \textrm{elpd}, we have $\Delta  \widehat{\mathrm{elpd}}^j = \max(\widehat{\mathrm{elpd}}) - \max(\widehat{\mathrm{elpd}}) = 0$. 
Therefore, for any other model $M_k$, the closer $\Delta  \widehat{\mathrm{elpd}}^k$ to zero, the better $M_k$ performs. 
When using $\Delta  \widehat{\mathrm{elpd}}$ to compare models, we also obtain estimated standard errors, denoted $\widehat{\text{se}}(\Delta \widehat{\textrm{elpd}})$.
We can identify a set of models with indistinguishable predictive performance compared to the best model as  
\begin{align}\label{eq:indist-models}
    \left\{ M_l: 0 \in \left[\Delta \widehat{\textrm{elpd}}^l \pm 2 \widehat{\text{se}}\left(\Delta \widehat{\textrm{elpd}}^l\right)\right] \right\}_{l=1, \cdots, L \leq K}.
\end{align}
\citet{sivula_uncertainty_2022} discuss the conditions when such a normal distribution approximation is justified and $\widehat{\text{se}}$ are reasonably calibrated. 
Specifically, $\widehat{\text{se}}$ can be underestimated in case of small data (say $n<100$) or when the models have very similar predictions. 
\citet{mclatchie_efficient_2023} justify why an ad hoc rule of $|\Delta\widehat{\mathrm{elpd}}|<4$ can be used to determine when the models have similar predictions and $\widehat{\text{se}}$ might be underestimated.
For iterative filtering, we are not interested in model selection based on optimal prediction, but instead want to reduce a set of models towards more promising candidates. 
To this end, we can use $\Delta\widehat{\mathrm{elpd}}$ and associated $\widehat{\text{se}}$ to identify models with largely inferior predictive abilities. 

\subsubsection*{Reliability of estimates obtained with PSIS-LOO-CV}\label{subsubsec:reliability-elpd}

When using importance weighting methods like PSIS \citep[][]{vehtari_pareto_2024} and data partitioning techniques like Bayesian LOO-CV to obtain estimates for evaluation criteria like \textrm{elpd}, it is essential to check the reliability of the obtained estimates.
In particular, since we want to filter by reliably assessed predictive performance, we diagnose estimates for \textrm{elpd} obtained with PSIS-LOO-CV by investigating the tail indices of the Pareto distribution used to smooth the importance sampling weights with the Pareto-$\hat k$ diagnostic \citep[][]{vehtari_pareto_2024}. 
Intuitively, reliable importance sampling is hard, if, for example, leaving out one observation affects the resulting posterior strongly. 
Following recommendations by \citet[]{vehtari_practical_2017} and \citet[]{vehtari_pareto_2024}, \textrm{elpd} estimates have to be considered unreliable if any of their associated Pareto-$\hat k$ values are larger than $0.7$. 

Given that we detect unreliable estimates for \textrm{elpd} obtained with PSIS-LOO-CV, different computational methods for computing leave-one-out predictive distributions can provide pathways towards trustworthy estimates.
For example, problems due to a few extreme observations can often be alleviated with moment matching in PSIS-LOO-CV as suggested by \citet[]{paananen_implicitly_2021} or estimates can be recomputed with K-fold-CV or brute-force LOO-CV\footnote{\citet[][]{paananen_implicitly_2021} demonstrate that brute-force LOO-CV can underestimate \textrm{elpd} in presence of outliers compared to analytically obtained \textrm{elpd}, leading to opposite bias compared to PSIS-LOO-CV in their experiments. 
Here, we do not assume access to an analytical solution and consider pessimistic estimates for \textrm{elpd} less costly for the modeller than unreliably overestimated predictive performance.} for problematic observations. 
Section \ref{subsec:epilepsy-comp} provides an example of problems in estimation of \textrm{elpd} when models include intercepts varying on the level of each observation.
We illustrate how to obtain reliable estimates in this scenario using a combination of default and integrated PSIS-LOO-CV, as well as re-computation with brute-force LOO-CV (see details in Appendix \ref{app:intloo}).

\subsection{Computation checks and iterative improvements}\label{subsec:computation-inference}

After having filtered out models with largely inferior predictive abilities, the remaining models enter the loop of ``Computation checks and iterative adjustments'' in Figure \ref{fig:approach}. 
Usually, posterior distributions in Bayesian data analysis tasks are too complex to be tractable, and we have to resort to numerical approximations to enable posterior inference. 
We focus on Markov chain Monte Carlo (MCMC) methods, in particular a variant of the dynamic Hamiltonian Monte Carlo no-U-turn sampler \citep[HMC-NUTS, ][]{hoffman_no-u-turn_2014}, as implemented in the probabilistic programming framework Stan \citep{carpenter_stan_2017, stan_development_team_stan_2023}. 
When using MCMC methods like HMC-NUTS, by definition, one can obtain draws from the target distribution when a Markov chain has reached the equilibrium \citep[see, e.g.,][]{neal_probabilistic_1993, brooks_handbook_2011}. 
Computation is not perfect and, in practice, we want to ensure sufficiently accurate sampling with Markov chains that get as close as possible to a common stationary distribution.

\subsubsection*{Diagnosing the sampling procedure}

Diagnostics for degrees of convergence and efficiency in sampling can provide important insights into the reliability of the obtained posterior draws when using MCMC methods like HMC-NUTS. 
In particular, comparing between- and within-chain variances with the improved potential scale reduction statistic $\hat{R}$ \citep[]{vehtari_rank-normalization_2021} or recent extensions like nested-$\hat{R}$ \citep[][]{margossian_nested_2024} provide tools to diagnose insufficiently mixed MCMC chains.

Measures of bulk and tail effective sample size (ESS) can be used to assess the efficiency of sampling for the given model in the bulk and the tails of the distribution, respectively \citep[see, e.g.,][]{geyer_practical_1992, brooks_handbook_2011, gelman_etal_bda3_2013, vehtari_rank-normalization_2021}. 
Specific to HMC-NUTS, divergent transitions can point out insufficient exploration of the posterior distribution \citep[see, e.g.,][]{leimkuhler_simulating_2005, gelman_etal_bda3_2013, betancourt_diagnosing_2016}. 
Section \ref{subsec:birthdays} illustrates iterative filtering with computational issues in inference indicated by divergent transitions, and provides an example of iteration towards an improved model parameterisation to alleviate issues in the sampling procedure.

\subsubsection*{Model complexity and computation}
For many real-world applications, it is sensible to assume that the underlying data generating process is complex and not directly accessible. 
A certain degree of experimenting and active assessment of computational issues for different models is likely required to successfully move from a statistical model to a working implementation in code \citep[]{gelman_bayesian_2020}. 
In this way, computational issues become an essential driver of the amount of necessary iteration, and iterative filtering with computation checks can direct limited resources to models that benefit most from improving computation.

\subsection{Iterating towards a better multiverse}\label{subsec:iterating}

As discussed in Section \ref{subsec:intro-workflow} and \ref{sec:support-iter} and illustrated in Section \ref{subsec:pred-eval-filtering} and \ref{subsec:computation-inference}, moving from modelling ideas to valid posterior inferences often involves iteration, for example, motivated by necessary adjustments or possibilities for improvement.
The suggested filtering approach reflects central aspects of Bayesian workflows described by \citet{gelman_bayesian_2020} like validating computation, addressing computational issues as well as evaluating and comparing models (see Figure 1 in \citet{gelman_bayesian_2020}).
In that way, iterative filtering provides access to essential modular components of the larger multi-objective modelling endeavour.

By design, iterative filtering identifies more promising candidate models from a larger set of models in the sense of a minimum viable Bayesian models described in Section \ref{subsec:filtering-min-viable}. 
At the same time, any filtered set of models is a preliminary result, and more iteration can lead to an improved filtered set. 
In particular, having defined and filtered an initial set of modelling choices, the modeller may be implored by the results observed to make changes to their hypotheses encapsulated in the initial models, or even consider new model structures. 
Our proposed framework enables the user to explicitly incorporate these discoveries through an iteration of the previous steps. 
That is, after having obtained a filtered set of models, the set can be extended and filtered again until a sufficient model or set of models is obtained.

Albeit sometimes unavoidable, modellers might not want to focus on one model until it fails, but instead find efficient, consistent and transparent transitions from an initial set of candidate models to more promising models.
Non-arbitrarily filtered models can offer a meaningful baseline for further analyses. 
If parallel exploration of multiple models is computationally costly, several simple models can be set up at once and filtered iteratively taking predictive abilities and computational issues into account.
Subsequently, the modeller can increase the complexity of the remaining models and filter again.
Section \ref{subsec:epilepsy-comp} provides an example of filtering an extension of a set of models from Section \ref{subsec:epilepsy-ppc}. 

\section{Case studies}\label{sec:case-studies}

In the following, we demonstrate how the suggested approach can simplify initial multiverses in real data examples where interest lies not only in sufficient predictive performance but also the effect of a treatment (Section \ref{subsec:epilepsy-ppc} and \ref{subsec:epilepsy-comp}) or specific day of the year (Section \ref{subsec:birthdays}) across different candidate models.  
Section \ref{subsec:epilepsy-ppc} presents filtering for predictive abilities for a set of models analysing the effect of an anticonvulsant therapy on epileptic seizure counts in patients with epilepsy. 
Subsequently, we extend the filtered set of models with additional modelling choices in Section \ref{subsec:epilepsy-comp} and demonstrate filtering for predictive abilities while taking into account the reliability of the estimated filtering criteria.
Section \ref{subsec:birthdays} illustrates how to iteratively filter a large set of models analysing the relative number of births on Halloween in the USA while detecting computational issues in inference and, if needed, adjusting the parameterisation for problematic models.
The code for the case studies is available at \url{https://github.com/annariha/multi-workflows}.

We use software tools for Bayesian subworkflows implemented in the \texttt{R} Statistical Software \citep[]{r_core_team_r_2023}{} for visualisations at different workflow steps (\texttt{bayesplot}, \citealp[]{gabry_bayesplot_2022}; \texttt{ggdist}, \citealp[]{kay_ggdist_2023}{}), working cleanly with posterior draws and predictions \citep[\texttt{posterior},][]{burkner_posterior_2023} as well as assessment of predictive performance and model comparisons \citep[\texttt{loo},][]{vehtari_loo_2020}.
Similar tools exist to some extent in other widely-used programming languages (see, e.g., \texttt{ArviZ}, \citealp[]{kumar_arviz_2019} with native packages in, e.g.,  \texttt{Python}, \citealp[]{python_software_foundation_python_2023} and \texttt{Julia}, \citealp[]{bezanson_julia_2017}).

\subsection{Epileptic seizure counts (Part I)}\label{subsec:epilepsy-ppc}
 
We use the data \texttt{brms::epilepsy} openly available in the \texttt{R}-package \texttt{brms} \citep{burkner_brms_2017}, initially published by \citet{leppik_controlled_1987} and previously analysed, for example, by \citet{thall_covariance_1990} and \citet{breslow_approximate_1993}, containing $236$ observations of a randomised trial of an anticonvulsant therapy for epilepsy patients with a treatment group that received the drug progabide $(N_t=124)$ and a control group $(N_c=112)$. 
All patients visited the clinic four times and the data contains information on seizure counts during and before the study as well as age of the patients. 

\begin{figure}[t]
    \centering
    \includegraphics[]{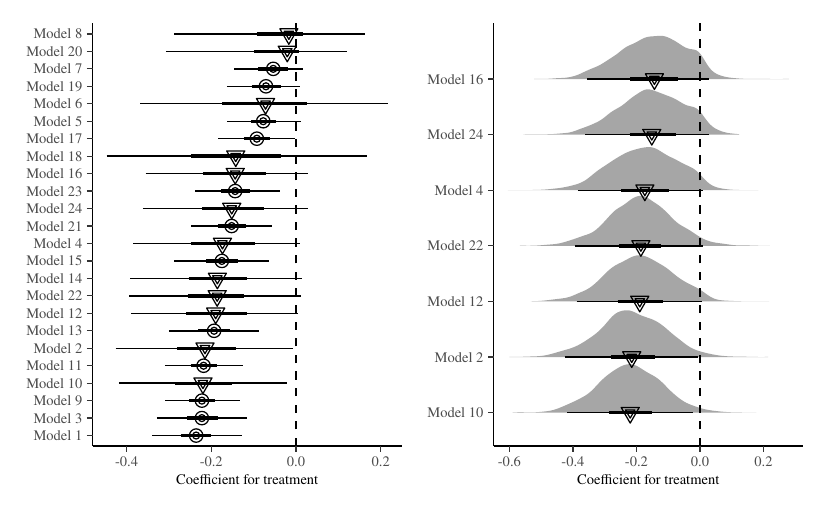}
    \vspace{-0.5cm}
    \caption{Epilepsy case study (Part I). Posterior results for the treatment coefficient with 50\% and 95\% posterior intervals ordered by posterior median value across the initial 24 models (left) and the remaining seven models after filtering (right). Initially, models differ considerably with regard to the conclusions about sign and range of the posterior coefficient of treatment. After filtering for predictive abilities, we observe more agreement within the filtered set of models.}
    \label{fig:epi-1-initial-posterior-trt-all-vs-indist}
\end{figure}

\paragraph{All considered models}
We combine different modelling choices to obtain the initial set of $24$ models summarised in Table \ref{tab:models-epi-1-ppc} in Appendix \ref{app:casestudy-epi-1-ppc}.  
To model the number of epileptic seizures, that is, non-negative counts, two immediate candidates for the distributional family of the observations are a Poisson and a negative Binomial distribution. 
Additionally, we consider different covariates and an interaction effect between treatment and baseline seizure count. 
Lastly, we compare default priors in \texttt{brms} \citep[][]{burkner_brms_2017}{}{} as well as a combination of default and regularised horseshoe (RHS) priors \citep[see][]{carvalho_handling_2009, carvalho_horseshoe_2010, piironen_sparsity_2017} with three degrees of freedom for the Student-$t$ prior of the local shrinkage parameters, using \texttt{brms::horseshoe(3)}.
The left subplot in Figure \ref{fig:epi-1-initial-posterior-trt-all-vs-indist} shows the median, and 50\% and 95\% posterior intervals for the treatment coefficient for each model. 
We see relatively high variation in conclusions that would be made based on these $24$ models. 
Choosing one model over any other affects the conclusions about the effect of treatment and, clearly, a joint consideration of models is a safer starting point for our modelling efforts than picking only one of the models. 

\begin{figure}[t]
    \centering
    \includegraphics{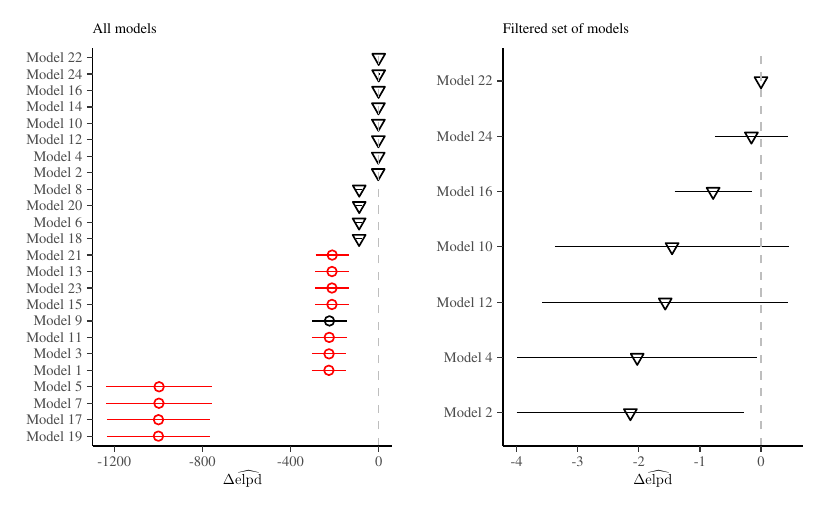}
    \vspace{-0.5cm}
    \caption{Epilepsy case study (Part I). Difference in expected log point-wise predictive density estimates ($\Delta \widehat{\textrm{elpd}}$) with estimated standard errors ($\widehat{\text{se}}$) ordered by mean $\Delta \widehat{\textrm{elpd}}$. Distributional family for the observations is either Poisson ($\circ$) or negative Binomial ($\triangledown$). The initial $24$ models are shown on the left, the remaining seven models on the right. Models with any Pareto-$\hat k$ values $>0.7$: \textcolor{red}{red}, no Pareto-$\hat k$'s $>0.7$: black.}
    \label{fig:epi-1-initial-elpddiff-2se-default}
\end{figure}

\paragraph{Filtering with predictive density estimates}
The left subplot in Figure \ref{fig:epi-1-initial-elpddiff-2se-default} shows $\Delta \widehat{\mathrm{elpd}}$ and $\widehat{\text{se}}(\Delta \widehat{\mathrm{elpd}})$ for all models and the right subplot shows only the seven models that are indistinguishable from the model with the highest estimated \textrm{elpd} according to their mean difference in \textrm{elpd} and the associated uncertainty $2 \cdot \widehat{\text{se}}(\Delta \widehat{\mathrm{elpd}})$. 
The right subplot in Figure \ref{fig:epi-1-initial-posterior-trt-all-vs-indist} shows posterior median, and 50\% and 95\% posterior intervals for the coefficient of treatment for the remaining seven models. 
We see that conclusions implied by the remaining models are similar. 

\begin{figure}[t]
    \centering
    \includegraphics{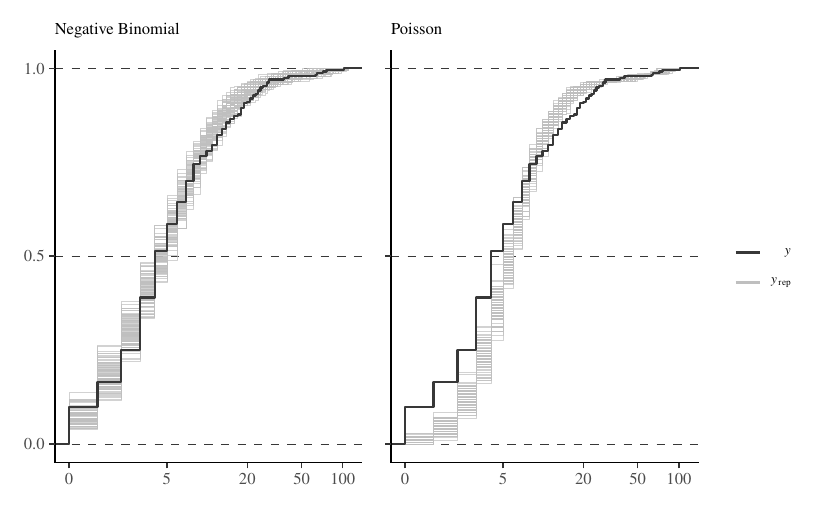}
    \vspace{-0.5cm}
    \caption{Epilepsy case study (Part I). Empirical cumulative density function (ECDF) estimates based on $100$ posterior replicates (\textcolor{gray}{grey} lines) and the distribution of the data (black line) illustrating the differences in expected and observed seizure counts for Model 22 (negative Binomial) and Model 21 (Poisson), differing only with respect to the chosen distributional family. Assuming a Poisson distribution seems to lead to more disagreement with the data compared to the model with negative Binomial distribution.}
    \label{fig:epi-1-initial-ecdf-model22-21}
\end{figure}

\paragraph{Filtering with posterior predictive checks} 
In the left subplot of Figure \ref{fig:epi-1-initial-elpddiff-2se-default}, \textrm{elpd} results where Pareto-$\hat k$ diagnostic indicated unreliable computation for PSIS-LOO-CV are highlighted with \textcolor{red}{red} colour.
Instead of using computationally more intensive CV approaches, we can use posterior predictive checking to rule out these models. 
For the given multiverse, all models with high Pareto-$\hat k$ assume a Poisson distribution as the distributional family  for the observations. 
Figure \ref{fig:epi-1-initial-ecdf-model22-21} shows posterior predictive checking results for the best performing model among models assuming a Poisson distribution (Model 21) and its counterpart that differs only with respect to the chosen distributional family for the observations (Model 22). 
The results suggest that the Poisson model is not an appropriate choice for this data. 
Figure \ref{fig:epi-1-initial-ppc-ecdf-model-17-24} in Appendix \ref{app:casestudy-epi-1-ppc} illustrates similar results of posterior predictive checks for other models, suggesting that there is no need to improve our CV methods for these models.
Figure \ref{fig:epi-1-initial-elpddiff-2se-intloo-reloo} in Appendix \ref{app:casestudy-epi-1-ppc} shows that in fact the filtering results would stay the same. 

\paragraph{A filtered set of models} 
With filtering, we are able to remove a large proportion of models that are clearly worse. 
The remaining set of seven filtered models can be analysed as usual in multiverse analysis. 

\subsection{Epileptic seizure counts (Part II)}\label{subsec:epilepsy-comp}

We extend the filtered set of seven models in Section \ref{subsec:epilepsy-ppc} to $175$ models.
Previously, with help of posterior predictive checking, we were able to avoid resolving computational issues in estimation of \textrm{elpd} for some models. 
With the extended set of models, iteration towards improved reliability of estimates for \textrm{elpd} is obligatory, as conclusions about the filtered set of models are affected.

\begin{figure}[t]
    \centering
    \includegraphics{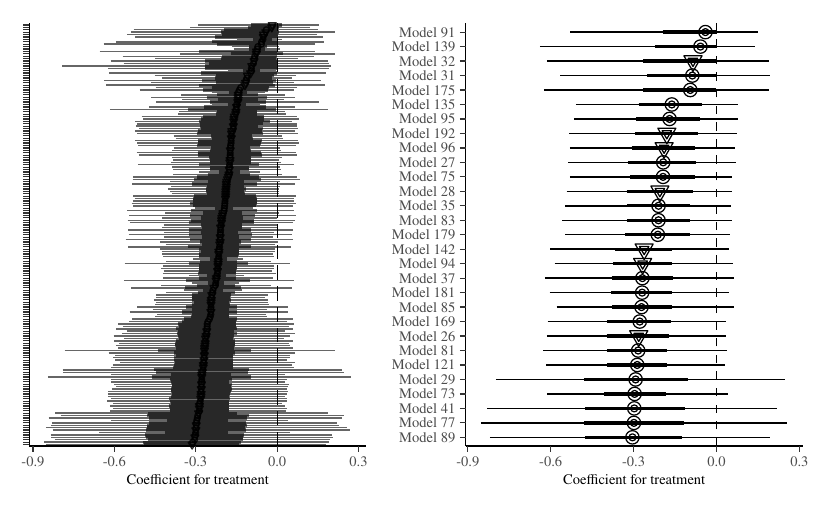}
    \vspace{-0.5cm}
    \caption{Epilepsy case study (Part II). Posterior results for the coefficient of treatment with 50\% and 95\% posterior intervals ordered by posterior median value across the $175$ models (left) and the filtered set of $29$ models (right). After filtering for predictive abilities, we observe more agreement within the filtered set of models.}
    \label{fig:epi-2-extended-posterior-trt-default-vs-filtered}
\end{figure}

\paragraph{Expanding the set of considered models} For a more detailed comparison of choosing a negative Binomial or a Poisson distribution under overdispersion, we add models with an additional effect varying by each observation.
Additionally, models can include additional variation on the level of each patient and each of the four visits and combinations thereof. 
Table \ref{tab:models-epi-2-psisloocv} in Appendix \ref{app:casestudy-epi-2-comp} summarises the resulting $175$ models. 
The left subplot in Figure \ref{fig:epi-2-extended-posterior-trt-default-vs-filtered} shows how the posterior 50\% and 95\% intervals and the posterior median for the coefficient of treatment vary across all $175$ models. 

\begin{figure}
    \centering
    \includegraphics{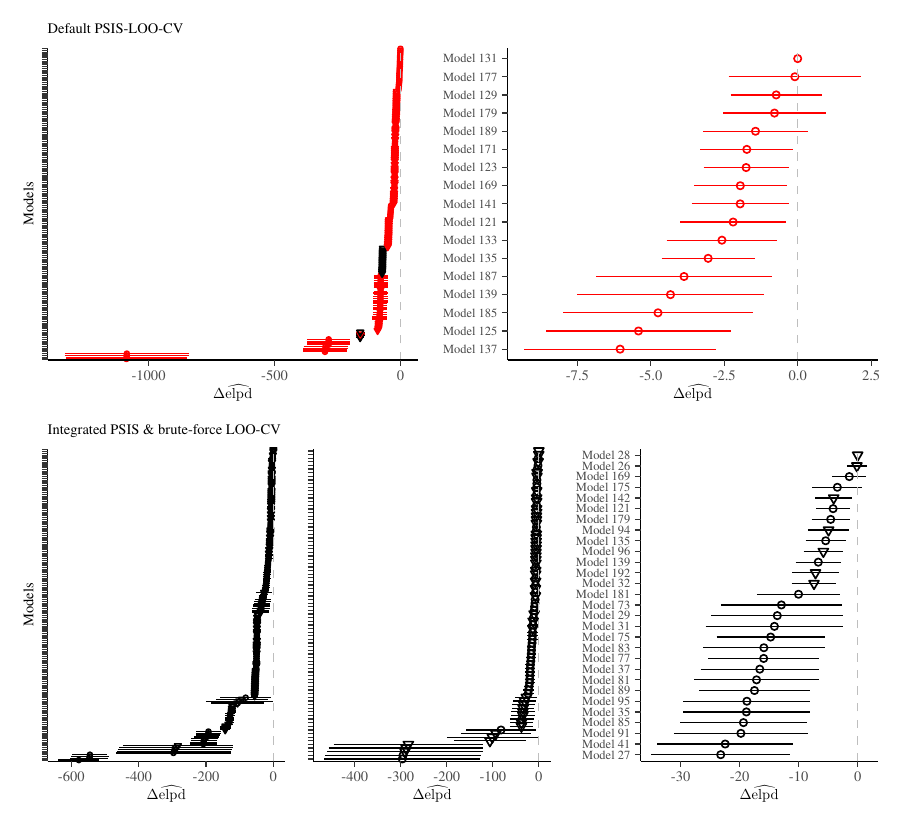}
    \vspace{-0.5cm}
    \caption{Epilepsy case study (Part II).
    Differences in expected log point-wise predictive density estimates ($\Delta \widehat{\textrm{elpd}}$) with $\widehat{\text{se}}$ ordered by mean $\Delta \widehat{\textrm{elpd}}$ for default PSIS-LOO-CV (top row) and integrated PSIS and brute-force LOO-CV (bottom row). 
    Distributional families for observations are Poisson ($\circ$) or negative Binomial ($\triangledown$), any Pareto-$\hat k$ values $>0.7$: \textcolor{red}{red}, no Pareto-$\hat k$'s $>0.7$: black.
    The two leftmost subplots show all $175$ models. 
    The top right subplot presents filtering with unreliable $\widehat{\textrm{elpd}}$, the bottom right subplot shows the filtered set of $29$ models after obtaining reliable estimates and checking the validity of the normal approximation.}
    \label{fig:epi-2-extended-elpddiff-2se-filtering-combined}
\end{figure}

\paragraph{Filtering with predictive performance}
We evaluate predictive performance of the models using
$\Delta \widehat{\mathrm{elpd}}$ and $\widehat{\text{se}}(\Delta \widehat{\mathrm{elpd}})$ (see Figure \ref{fig:epi-2-extended-elpddiff-2se-filtering-combined}).
In contrast to the results presented in Section \ref{subsec:epilepsy-ppc}, the top right subplot in Figure \ref{fig:epi-2-extended-elpddiff-2se-filtering-combined} shows that direct filtering for models indistinguishable with respect to $\Delta \widehat{\mathrm{elpd}}$ and $2 \cdot \widehat{\text{se}}(\Delta \widehat{\mathrm{elpd}})$ would focus exclusively on models with unreliably estimated \textrm{elpd}. 
As discussed in Section \ref{subsec:pred-eval-filtering}, any conclusions about promising candidate models should be based on a reliable assessment of predictive performance.
This is an example of changing and checking in Figure \ref{fig:approach}, since computation of estimates of \textrm{elpd} needs to be adjusted to resolve the detected issues and obtain reliably estimated filtering criteria before further filtering attempts.  

\paragraph{Computational issues in estimation of \textrm{elpd}}
Investigation of the modelling choices underlines that issues in reliability of estimates of \textrm{elpd} are mostly related to models including additional variation on the level of each observation. 
To alleviate challenges of importance sampling and, in turn, PSIS-LOO-CV, for models with an effect varying by each observation, we can separate the varying intercept from the rest of the model and integrate over the observation-level intercepts.
This approach is called integrated PSIS-LOO-CV. 
Appendix \ref{app:intloo} explains the procedure in more detail. 
After obtaining estimates with integrated PSIS-LOO-CV, only a few high ($> 0.7$) Pareto-$\hat{k}$ values remain for some models for which estimates for \textrm{elpd} are recomputed with brute-force LOO-CV using \texttt{brms::reloo()} from the \texttt{brms} package \citep[]{burkner_brms_2017}.
Comparing results before and after modifying computation in the top and bottom left subplots in Figure \ref{fig:epi-2-extended-elpddiff-2se-filtering-combined}, we observe lower mean absolute differences in $\widehat{\mathrm{elpd}}$ and larger associated standard errors $\widehat{\text{se}}(\Delta \widehat{\mathrm{elpd}})$ for most models after obtaining reliable estimates for all models. 

\paragraph{Filtering with reliably estimated \textrm{elpd}}
As illustrated in the top right and bottom central subplots in Figure \ref{fig:epi-2-extended-elpddiff-2se-filtering-combined}, improving reliability of estimates for \textrm{elpd} with integrated PSIS-LOO-CV and brute-force LOO-CV substantially changes the set of filtered models.

\paragraph{Checking validity of the normal approximation when summarising point-wise estimated \textrm{elpd}}
After improving computation, the bottom central subplot in Figure \ref{fig:epi-2-extended-elpddiff-2se-filtering-combined} shows that some models have large absolute $\Delta \widehat{\mathrm{elpd}}$ but also large estimated standard errors $\widehat{\text{se}}(\Delta \widehat{\mathrm{elpd}})$. 
It seems unlikely that these models are actually promising candidates. 
As touched upon in Section \ref{subsec:pred-eval-filtering}, summarising expected log point-wise predictive densities with $\Delta \widehat{\mathrm{elpd}}$ and associated standard errors $\widehat{\text{se}}(\Delta \widehat{\mathrm{elpd}})$ relies on approximating the distribution of point-wise differences in estimated \textrm{elpd} with a normal distribution. 
Large standard errors can indicate issues with the validity of this approximation. 
To investigate whether results for all $86$ models are based on valid approximations, we diagnose point-wise differences in estimated elpd by comparing the Pareto-$\hat{k}$ values to the Pareto-$\hat{k}$ threshold and the minimum required sample size for valid approximation with a normal distribution \citep{vehtari_pareto_2024} using \texttt{posterior::pareto\_diags()} from the \texttt{posterior} package \citep[]{burkner_posterior_2023}{}. 
The Pareto-$\hat{k}$ value for the point-wise differences is larger than the sample size specific threshold (here: $0.58$) and the required minimum sample size exceeds the number of available observations (here: $236$) for several models for which, consequently, the filtering using $\Delta \widehat{\mathrm{elpd}}$ and $2 \cdot \widehat{\text{se}}(\Delta \widehat{\mathrm{elpd}})$ is not accurate. 

\paragraph{Predictive abilities with few extreme observations}
Investigating the point-wise values for \textrm{elpd} and differences in \textrm{elpd} reveals that predictive performance for some models is largely affected by a few extreme observations, which translates to large standard errors and failure of the normal approximation.
Checking for extreme observations, we identify patient $25$ and patient $49$ with magnitude or pattern of seizure counts that differ largely from the rest of the patients. 
Indeed, extreme observations for these patients largely correspond to the points where the problematic models perform a lot worse than other models, which then translates to large estimated standard errors of the mean difference in \textrm{elpd} (see examples in Figure \ref{fig:epi-2-pointwise-elpddiff-worst-models} in Appendix \ref{app:casestudy-epi-2-comp}).
In absence of expert knowledge that would allow us to identify data entry or measurement errors in these cases, we consult previous research using this data set \citep[see, e.g.,][]{thall_covariance_1990, breslow_approximate_1993}{}{}. 
In particular, \citet{thall_covariance_1990} advise that for extreme observations in this data set ``deletion [...] has no clinical basis'' \citep[][p.666]{thall_covariance_1990}{}{}.
Therefore, given the available data, our filtered set of models should only include models that are able to model the extreme observations in the data set sufficiently well. 

\paragraph{A filtered set of models} Out of $175$ initial models, we finally obtain $29$ models. 
Given a larger set of more complex models, iteration towards improved reliability of estimates for \textrm{elpd} is obligatory when conclusions about the filtered set of models are affected.
The full model specifications for the remaining $29$ models are highlighted in \mybox[fill=gray!50]{grey} in Table \ref{tab:models-epi-2-psisloocv} in Appendix \ref{app:casestudy-epi-2-comp}. 
The right subplot of Figure \ref{fig:epi-2-extended-posterior-trt-default-vs-filtered} shows posterior median, and 50\% and 95\% posterior intervals for the filtered set of models. 
Now, tools from multiverse analysis can be used for this reduced set of models. 

\paragraph{Identifying structure in the filtered set of models} In the set of $29$ models, we can identify the most and least complex models, as well as models that are nested within each other. 
In particular, the most complex model assuming a negative Binomial distribution as the distributional family for the observations (Model 192 in \ref{tab:models-epi-2-psisloocv}) includes all covariates and group-level effects on the level of each observation, each patient and each visit and assumes a combination of default and regularised horseshoe priors.
Moreover, the most complex model assuming a Poisson distribution (Model 181 in Table \ref{tab:models-epi-2-psisloocv}) contains all covariates as well as an interaction effect between treatment and baseline seizure count and group-level effects on the level of each observation, each patient and each visit and assumes default prior settings in \texttt{brms::brm()}. 
The least complex model in the set of $29$ models uses a Poisson distribution as the distributional family for the observations, includes a covariate indicating whether the patient received treatment and a group-level effect for each patient and assumes a combination of default and regularised horseshoe priors (see Model 31 in Table \ref{tab:models-epi-2-psisloocv}).

\subsection{Birthdays (Reparameterisation)}\label{subsec:birthdays}
After giving an example for tending to computational issues in filtering criteria, we now present a scenario in which difficult posterior geometry can cause computational issues in inference when analysing the relative number of daily births in the USA from $1969$ to $1988$. 
This requires the modeller to combine assessment of predictive performance with iterative modification of models via reparameterisation until reliable sampling is possible. 
We use natality data from the National Vital Statistics System provided by Google BigQuery and exported by Chris Mulligan and Robert Kern containing $7305$ observations of the number of births with the corresponding day, month, and year as well as day of the year and day of the week in the USA from $01.01.1969$ to $31.12.1988$. 

\paragraph{All considered models}
We model the registered number of daily births with models of varying complexity, ranging from a normal hierarchical model for the day of the year effect to a model which combines Gaussian process (GP) priors \citep[see, e.g.,][]{rasmussen_gaussian_2005} on multiple unknown time dependent functions with a regularised horseshoe (RHS) prior \citep[see][]{carvalho_handling_2009, carvalho_horseshoe_2010, piironen_sparsity_2017} for the day of the year effect. 
The modelling choices are inspired by a subset of the components of the model showcased by \citet[][Chapter $21$]{gelman_etal_bda3_2013}. 
In total, we consider $144$ different models, constructed by individually switching out model components. 
Our most complex model combines a GP, a yearly periodic GP, a GP controlling the strength of a day of the week effect, a floating holiday effect and a day of the year effect with an RHS prior. 
Table \ref{tab:models-birthdays-reparam} in Appendix \ref{appsubsec:modelspecs-casestudy-birthdays} summarises the model components and the resulting models.

\begin{figure}[t]
    \centering
    \includegraphics[]{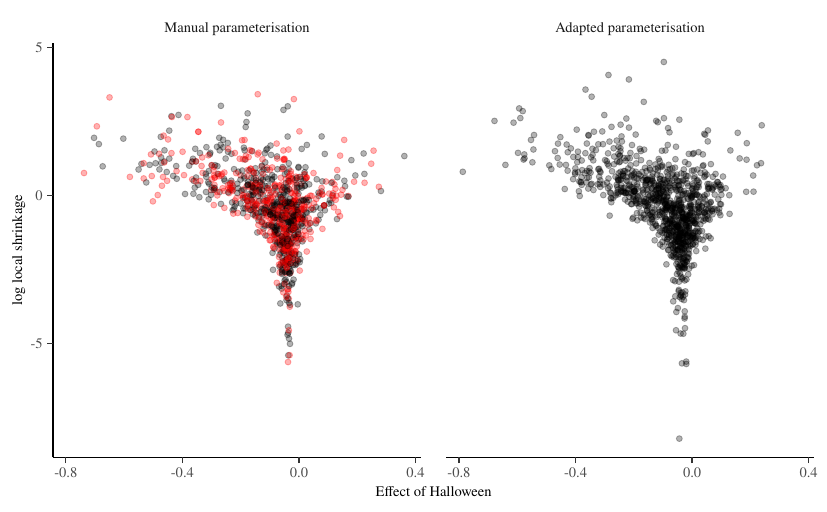}
    \vspace{-0.3cm}
    \caption{Birthdays case study (Reparameterisation). Posterior draws for the Halloween effect and log local shrinkage parameter of the RHS prior on day-of-the-year effect for the simplest model containing the RHS model component (Model 6 in Table \ref{tab:models-birthdays-reparam}), grouped by parameterisation (left: manual, right: adapted) and diverged transitions (\textcolor{red}{red}: yes, black: no).  
    Slightly more than 20\% of transitions for the manual parameterisation ended in divergences, only one did so for the adapted parameterisation. The average number of leapfrog steps was $913$ for the manual and $127$ for the adapted parameterisation.}
    \label{fig:birthdays-funnel-before-after}
\end{figure}

\paragraph{Computational issues in inference}
The high number of observations ($N=7305$) and the nonlinear influence of the unknown time dependent functions makes exact covariance matrix based GP implementations computationally slow. 
We instead use a Hilbert space basis function based approximation \citep{solin_hilbert_2020}, which has come to be known as Hilbert space basis function GP \citep[HSGP, see][]{riutort-mayol_practical_2022}. 
However, these HSGPs as well as the RHS prior come with their own computational challenges. 
Both lead to model parameter priors with scales that depend strongly on other model parameters, similar to the situation exemplified in Neal’s funnel \citep{neal_slice_2003}.  
Neither monolithic centring nor monolithic non-centring removes these dependencies due to the effect of the likelihood.
For many of the considered models, this leads to a significant number of divergences when sampling, ranging up to almost 50\% divergent transitions (see Figure \ref{fig:birthdays-funnel-before-after}). 
If the number of divergent transitions is zero, we will assume that sampling has worked if the other convergence diagnostics indicate so. 
However, even a small fraction of divergent transitions can mean that the sampler has failed to explore parts of the posterior.

\paragraph{Modifying parameterisation to improve sampling}
For models for which we encounter divergences during sampling, we use the obtained draws to estimate a model parameterisation which minimises the Kullback–Leibler divergence between the posterior distribution and a multivariate normal distribution. 
As demonstrated by \citet{gorinova_automatic_2020}, such a reparameterisation tends to improve the posterior geometry and facilitate sampling.
This inferred reparameterisation together with an increased target acceptance rate allows us to sample even from the challenging posteriors with no or only very few divergent transitions. 

\begin{figure}[t]
    \centering
    \includegraphics{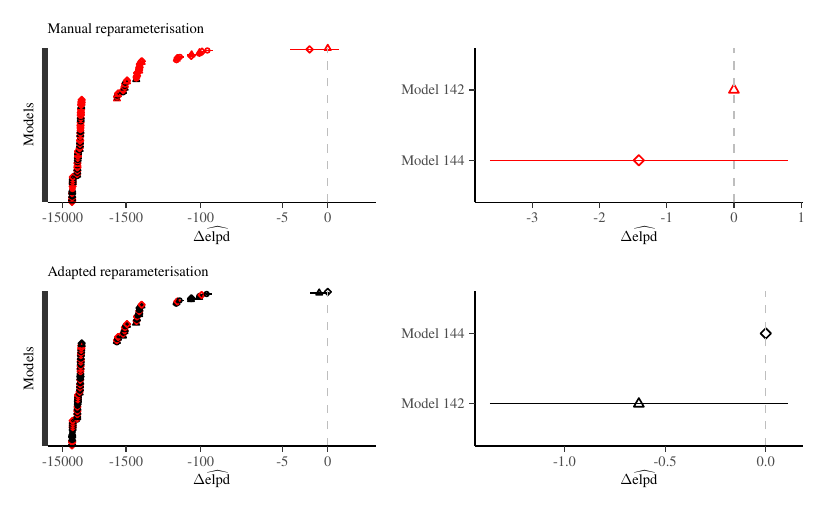}
    \vspace{-0.5cm}
    \caption{Birthdays case study (Reparameterisation).  Differences in expected log point-wise predictive density estimates ($\Delta \widehat{\textrm{elpd}}$) with $\widehat{\text{se}}$ ordered by mean $\Delta \widehat{\textrm{elpd}}$, by parameterisation (row), day-of-the-year family ($\circ$: normal, $\triangle$: Student's t, $\Diamond$: RHS) and occurred divergences (\textcolor{red}{red}: yes, black: no).}
    \label{fig:birthdays-elpddiff-combined}
\end{figure}

\paragraph{Predictive model evaluation}
As before, to establish a ranking in the set of models and to investigate whether we have to further improve sampling, we compare predictive performance between the models using estimated \textrm{elpd} (see Figure \ref{fig:birthdays-elpddiff-combined}). 
This time, the reliability of the \textrm{elpd} estimates is affected strongly by the reliability of the sampling, but not at all by the reliability of the PSIS-LOO-CV. 
As in Section \ref{subsec:epilepsy-comp}, it is not only the lowest ranking models where the reliability of the elpd estimates is in question, but it is in particular the highest ranking and most complex models where sampling seems to have been problematic. 
After improving sampling, the ranking of the two best models flips and all diagnostics indicate that estimation was successful for the important models. 

\begin{figure}[t]
    \centering 
    \includegraphics{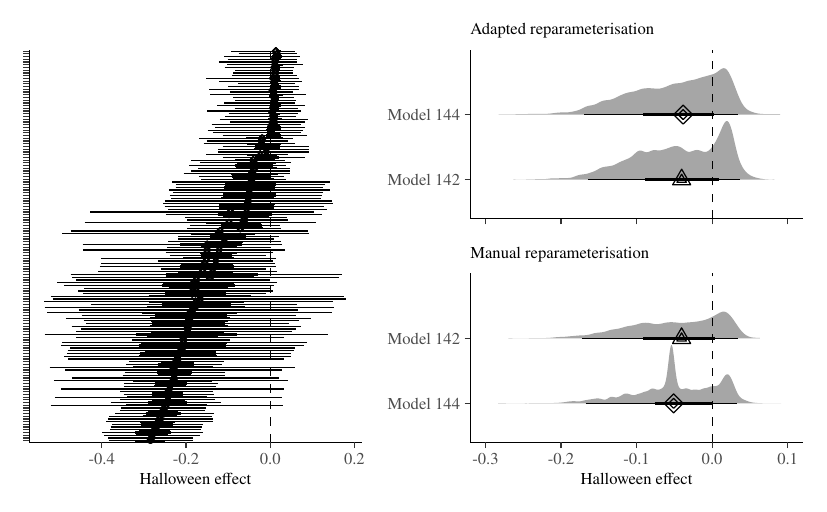}
    \vspace{-0.6cm}
    \caption{Birthdays case study (Reparameterisation). Effect of Halloween on registered births with 50\% and 95\% probability intervals. The left subplot reports results for all $144$ models, the right subplot shows the remaining two models grouped row-wise by manual and adapted parameterisation, respectively.}
    \label{fig:birthdays-qoi-halloween}
\end{figure}

\paragraph{A filtered set of models}
Following the original analysis by \citet{levy_influence_2011}, we look at the effect of Halloween on the number of registered births (see Figure \ref{fig:birthdays-qoi-halloween}). 
As in Section \ref{subsec:epilepsy-ppc} an unfiltered multiverse of models would have suggested considerable variation in plausible effect sizes. 
By filtering out models with very bad predictive performance and iteratively improving computation for the good model candidates, we are able to narrow down the set of $144$ models to just two candidate models, which both give similar predictions and lead to the same conclusions about the effect in question. 
As in Section \ref{subsec:epilepsy-ppc}, but unlike in Section \ref{subsec:epilepsy-comp}, the set of candidate models does not change, even though their internal ranking changes. 
However, as in Section \ref{subsec:epilepsy-comp}, correcting the computational issues does not notably affect the conclusions about the effect in question.

\paragraph{The filtered set of models as the new baseline}
While our procedure has left us with two models that we can tentatively accept, there is no reason to blindly trust their predictions or their inference. 
Our procedure placing the most complex models at the top should instead encourage us to consider these models as the new baseline, from which we can continue to explore the model space, if needed or wanted. 
If we choose to do so, we can and should expect to find models which capture the underlying process considerably or even just slightly better, see, for example, the inspiration for our model components by \cite{gelman_etal_bda3_2013}. 
These models may or may not change the conclusions about the effect of interest, and as such we should accept any conclusions only tentatively, until new information is added.

\section{Discussion}\label{sec:discussion}
Iterative filtering for multiverse analysis can identify promising candidates in large sets of models, given that causal constraints have been taken into account.
Tending to computational issues in tandem with filtering for predictive abilities enables pruning a large set of models to a manageable amount, for example, from $24$ to seven models in Section \ref{subsec:epilepsy-ppc}, from $175$ to $29$ models in Section \ref{subsec:epilepsy-comp} and from $144$ to two models in Section \ref{subsec:birthdays}.
Naturally, this work is not without limitations. 
The task of supporting iterative Bayesian modelling workflows is multi-faceted and raises questions outside the scope of this paper.

\subsection{Challenges and limitations of this work}

Our filtering approach focuses on predictive abilities, investigation of posterior results of the quantity of interest and computation checks as aspects of central importance. 
\citet{wang_difficulty_2015} illustrate that predictive accuracy alone can be an insufficient criterion for model comparison.
Similarities in predictive performance are not necessarily an indicator for similarity in the posteriors of the quantity of interest. 
After filtering based on predictive performance, multiverse analysis provides means for assessing possible differences in interesting posterior quantities.

We have assumed that no causally nonsensical models are included in the set of candidate models.
If causal constraints are not taken into account, filtering may remove causally sensible models with worse predictive performance than causally nonsensical models with better predictive performance \citep[for an example of predictive model selection favouring a causally wrong model see, e.g., Section 7.5.1 by][]{mcelreath_rethinking_2020}.

Even though the proposed approach enables modellers to jointly investigate multiple models with less overhead, it is clear that with more time, and more iterations, we can always obtain an even larger set of models \citep[see, e.g.,][]{bernardo_bayesian_1994, simonsohn_specification_2020}. 
As of now, iterative filtering does not provide an automatic tool, but requires activity and engagement from the modeller. 
Creating and investigating sets of models is subject to computational and cognitive constraints, limitations in modelling ideas, as well as availability of data and time. 
If the modeller is faced with an extremely large amount of models, resource limitations can make it infeasible to consider all models at once and the proposed approach can fail to reduce the set of models to a manageable amount. 

We do not compare different possible multiverses, and instead assume a fixed set of modelling choices that give rise to one multiverse of models in the examples presented in Section \ref{sec:case-studies}. 
As \citet{bell_modeling_2022} point out, it is to be expected that the choices required for any multiverse analysis might in turn affect the results: ``choice of search space is itself a choice'' \citep[][p. 8]{bell_modeling_2022}.
This is also related to ``model list selection'' \citep[][p. 30]{clarke_cheat_2023}, that is, the question of how to select appropriate lists of models to compare that sufficiently represent the underlying but inaccessible entirety of possible models, which \citet{clarke_cheat_2023} identify as one of the pressing issues for future research on assessment of predictive abilities of several candidate models in a Bayesian framework.

\subsection{Outlook}
After having obtained an iteratively filtered set of models, it remains to be explored how the modeller continues; especially if interest lies in decision-making based on the filtered set of models. 
Iterative filtering focuses on removing clearly inferior models from the multiverse, and not identifying a single best model.
If a good reference model is available and interest lies predominantly in predictions, projection predictive inference could provide a systematic way to identify a simpler model that matches the predictive performance of the reference model from a set of candidates \citep[see, e.g.,][]{pavone_using_2023, piironen_projective_2020, piironen_projpred_2023, mclatchie_robust_2023}.  
Another possibility is model averaging for the remaining models, for example, via model stacking \citep[]{yao_using_2018}{}, since the joint inferences are then based on a pruned set excluding largely inferior models. 

Iterative filtering for multiverse analysis is an attempt to make required and desired iteration in a Bayesian modelling workflow more tangible. 
When building Bayesian models, modellers routinely face the tasks of evaluating computation as well as assessing predictive and inferential abilities for multiple models.
To further support Bayesian data analysis tasks, it is important to develop more approaches that  
\begin{enumerate}[nosep]
    \item[\textbullet] allow joint consideration of multiple models;
    \item[\textbullet] implement connections between different subtasks in Bayesian workflows;
    \item[\textbullet] and support required and desired iteration throughout.
\end{enumerate}
The goal of supporting Bayesian workflows and, more specifically, the iterative aspects therein, is connected to a larger discussion about vital properties of useful modelling workflows.
\citet[]{broderick_toward_2023}{} emphasise that trust in modelling practices is linked to the robustness of conclusions to different underlying assumptions. 
Every model can be seen as a hypothesis of the modeller, formulated by a combination of modelling choices \citep[][]{gelman_philosophy_2012}. 
By acknowledging the existence of several such hypotheses, trustworthy Bayesian modelling workflows need to allow the consistent and transparent consideration of several candidate models.  

\subsection*{Acknowledgements}

We thank Andrew Gelman and Seth Axen for helpful comments. 
We acknowledge the computational resources provided by the Aalto Science-IT project. 
This work was supported by the Research Council of Finland Flagship programme: Finnish Center for Artificial Intelligence (FCAI),
and Research Council of Finland project “Safe iterative model building” (340721).
We also thank our colleagues who provided helpful feedback, particularly Andrew R. Johnson, Yann McLatchie, Frank Weber, and Noa Kallioinen. 

\bibliography{references}

\newpage

\begin{appendices}

\section{Some suggested requirements for consistent and efficient iterative modelling}\label{app:requirements-iteration}

\begin{figure}[hbp]
    \centering
    \includegraphics{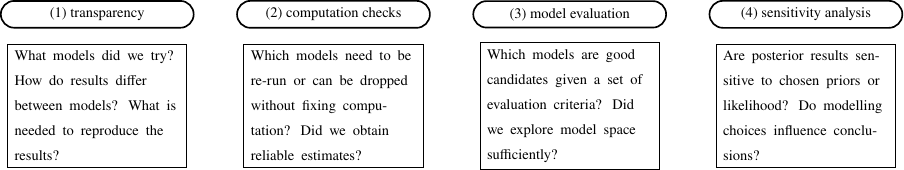}
    \caption{Overview of some suggested requirements for consistent and efficient iteration in Bayesian modelling workflows and examples of questions a modeller might ask in reflection of these requirements.}
    \label{fig:overview-requirements-iteration}
\end{figure}

\section{Integrated PSIS-LOO-CV}\label{app:intloo}

For a model with an intercept varying on the level of each observation, it is possible to compute the log likelihood by separating the varying intercept $r_i$ from the rest of the model and integrating over each $r_i$ which leads to 
\begin{align} \label{eq:intloolikelihood}
    \log p(y_i \mid X_i, \theta^s_{-r_i}) = \log \int_{-\infty}^{\infty} p(y_i \mid X_i, r_i, \theta^s_{-r_i}) p(r_i \mid \theta^s) d r_i ,
\end{align}
for each observation $i=1,\cdots, N$ and each posterior draw $s=1, \cdots, S$. 
For the $s$-th posterior draw, $\theta^s_{-r_i}$ denotes the vector of model parameters excluding the intercept varying by observations $r_i$ for the $i$-th observation. 
Using PSIS-LOO-CV means that we are using importance sampling, for example, with the posterior density $p(\theta \mid Y)$ as the proposal and $p(\theta \mid Y_{-i})$ as the target densities for each observation in the leave-one-out approach. 
Since the posterior density $p(\theta \mid Y)$ includes the marginal effect of each observation-level intercept $r_i$, leaving out one observation at a time means that we will end up comparing the marginal $p(r_i \mid Y)$ given all observations of the outcome with $p(r_i \mid Y_{-i})$ with observation $i$ removed. 
Clearly, the difference between these marginals will be big, when we are removing the $i$-th observation from the marginal for $r_i$, whereas, for example, removing the $j$-th observation will have a smaller effect on the marginal of $r_i$ but again a big effect on the marginal of $r_j$ and so on. 
This means that 
\begin{align}
    \| p(r_i | Y)  -  p(r_i | Y_{-i}) \| \gg 0,
\end{align}
and we observe that this translates to the proposal density $p(\theta | Y)$ and the target density $p(\theta | Y_{-i}) $ as well since they contain the marginals $p(r_i | Y)$ and $p(r_i | Y_{-i})$, respectively. 

Since the marginal posterior of the intercept varying on the level of each observation, denoted $r_i$,  differs largely between each observation while the rest of the posterior is not affected, we can separate the intercept from the rest of the model and integrate over the intercepts that vary by observations. 
As detailed in equation \ref{eq:intloolikelihood}, this allows us to incorporate $r_i$ into our proposal and target distribution without having to consider $p(r_i \mid Y)$ and $p(r_i \mid Y_{-i})$ separately by integrating over $r_i$ for each observation $i$.
Notably, due to the influence of the marginal of $r_i$, we can assume that the proposal posterior density $p(\theta | Y)$ differs more from the target $p(\theta | Y_{-i})$ than the posterior density $p(\theta_{r_i} \mid Y)$ and the corresponding target $p(\theta_{r_i} | Y_i)$. In particular, 
\begin{align}
    \| p(\theta \mid Y) - p(\theta \mid Y_{-i}) \| \gg 
    \| p(\theta_{r_i} \mid Y) - p(\theta_{r_i} \mid Y_{-i}) \| .
\end{align}
Intuitively, one would expect that this relationship makes it easier to obtain reliable importance sampling results due to a thinner tail of the distribution of importance sampling weights when the proposal and target distribution are closer to each other.
With the log likelihood evaluations obtained by integrating over the intercepts varying on the level of each observation we remove the direct influence of each individual observation and can then again use PSIS-LOO-CV to obtain reliable \textrm{elpd} estimates as exemplified in Section \ref{subsec:epilepsy-comp}. 

\section{Epilepsy case study (Part I)}\label{app:casestudy-epi-1-ppc}

\subsection{Modelling choices}\label{appsubsec:models-casestudy-epi-1-ppc}
{\small 
\begin{longtable}[c]{llll}
    \caption{Epilepsy case study (Part I). All modelling choices, corresponding Model ID's and model formulae used to fit the multiverse of $24$ models with \texttt{brms::brm()}. Rows highlighted in \mybox[fill=gray!50]{grey} indicate the filtered set of seven models depicted in the right subplot in Figure \ref{fig:epi-1-initial-elpddiff-2se-default} and \ref{fig:epi-1-initial-elpddiff-2se-intloo-reloo}.} 
    \label{tab:models-epi-1-ppc}\\
    \toprule
    Model ID & Obs. family & Priors & Formula\\
    \midrule
    \endfirsthead
    \multicolumn{4}{@{}l}{\textit{(continued)}}\\
    \toprule
    Model ID & Obs. family & Priors & Formula\\
    \midrule
    \endhead

    \endfoot
    \bottomrule
    \endlastfoot
    Model 1 & Poisson & default in \texttt{brms} & \texttt{count $\sim$ zBase * Trt}\\
    \rowcolor{lightgray} Model 2 & Neg. Binomial & default in \texttt{brms} & \texttt{count $\sim$ zBase * Trt}\\
    Model 3 & Poisson & \texttt{brms::horseshoe(3)} & \texttt{count $\sim$ zBase * Trt}\\
    \rowcolor{lightgray} Model 4 & Neg. Binomial & \texttt{brms::horseshoe(3)} & \texttt{count $\sim$ zBase * Trt}\\
    Model 5 & Poisson & default in \texttt{brms} & \texttt{count $\sim$ Trt}\\
    Model 6 & Neg. Binomial & default in \texttt{brms} & \texttt{count $\sim$ Trt}\\
    Model 7 & Poisson & \texttt{brms::horseshoe(3)} & \texttt{count $\sim$ Trt}\\
    Model 8 & Neg. Binomial & \texttt{brms::horseshoe(3)} & \texttt{count $\sim$ Trt}\\
    Model 9 & Poisson & default in \texttt{brms} & \texttt{count $\sim$ Trt+zBase}\\
    \rowcolor{lightgray} Model 10 & Neg. Binomial & default in \texttt{brms} & \texttt{count $\sim$ Trt+zBase}\\
    Model 11 & Poisson & \texttt{brms::horseshoe(3)} & \texttt{count $\sim$ Trt+zBase}\\
    \rowcolor{lightgray} Model 12 & Neg. Binomial & \texttt{brms::horseshoe(3)} & \texttt{count $\sim$ Trt+zBase}\\
    Model 13 & Poisson & default in \texttt{brms} & \texttt{count $\sim$ zBase * Trt+zAge}\\
    Model 14 & Neg. Binomial & default in \texttt{brms} & \texttt{count $\sim$ zBase * Trt+zAge}\\
    Model 15 & Poisson & \texttt{brms::horseshoe(3)} & \texttt{count $\sim$ zBase * Trt+zAge}\\
    \rowcolor{lightgray} Model 16 & Neg. Binomial & \texttt{brms::horseshoe(3)} & \texttt{count $\sim$ zBase * Trt+zAge}\\
    Model 17 & Poisson & default in \texttt{brms} & \texttt{count $\sim$ Trt+zAge}\\
    Model 18 & Neg. Binomial & default in \texttt{brms} & \texttt{count $\sim$ Trt+zAge}\\
    Model 19 & Poisson & \texttt{brms::horseshoe(3)} & \texttt{count $\sim$ Trt+zAge}\\
    Model 20 & Neg. Binomial & \texttt{brms::horseshoe(3)} & \texttt{count $\sim$ Trt+zAge}\\
    Model 21 & Poisson & default in \texttt{brms} & \texttt{count $\sim$ Trt+zBase+zAge}\\
    \rowcolor{lightgray} Model 22 & Neg. Binomial & default in \texttt{brms} & \texttt{count $\sim$ Trt+zBase+zAge}\\
    Model 23 & Poisson & \texttt{brms::horseshoe(3)} & \texttt{count $\sim$ Trt+zBase+zAge}\\
    \rowcolor{lightgray} Model 24 & Neg. Binomial & \texttt{brms::horseshoe(3)} & \texttt{count $\sim$ Trt+zBase+zAge}\\*
\end{longtable}
}    

\clearpage

\subsection{Improving estimation of \textrm{elpd} does not affect filtered set of models}\label{appsubsec:initial-elpddiff-2se-intloo-reloo-casestudy-epi-1-ppc}

\begin{figure}[hbp]
    \centering
    \includegraphics{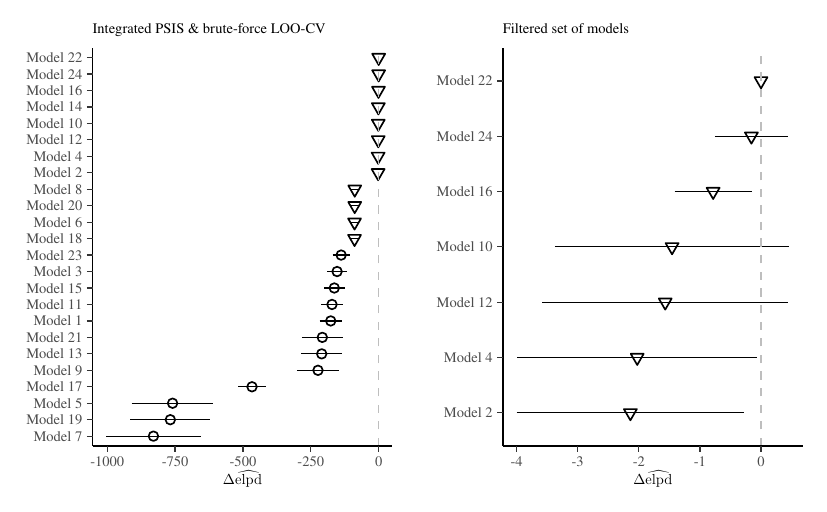}
    \vspace{-0.5cm}
    \caption{Epilepsy case study (Part I). Difference in expected log point-wise predictive density estimates ($\Delta \widehat{\textrm{elpd}}$) with estimated standard errors ($\widehat{\text{se}}$) ordered by mean $\Delta \widehat{\textrm{elpd}}$. Results for the initial $24$ models using integrated PSIS-LOO-CV and brute-force LOO-CV (left) and the remaining seven models (right). The filtered set of models is not affected by improving estimation of \textrm{elpd}.}
    \label{fig:epi-1-initial-elpddiff-2se-intloo-reloo}
\end{figure}

\clearpage

\subsection{ECDF plots for posterior predictive checks}\label{appsubsec:ecdf-plots-casestudy-epi-1-ppc}

\begin{figure}[hbp]
    \centering
    \includegraphics{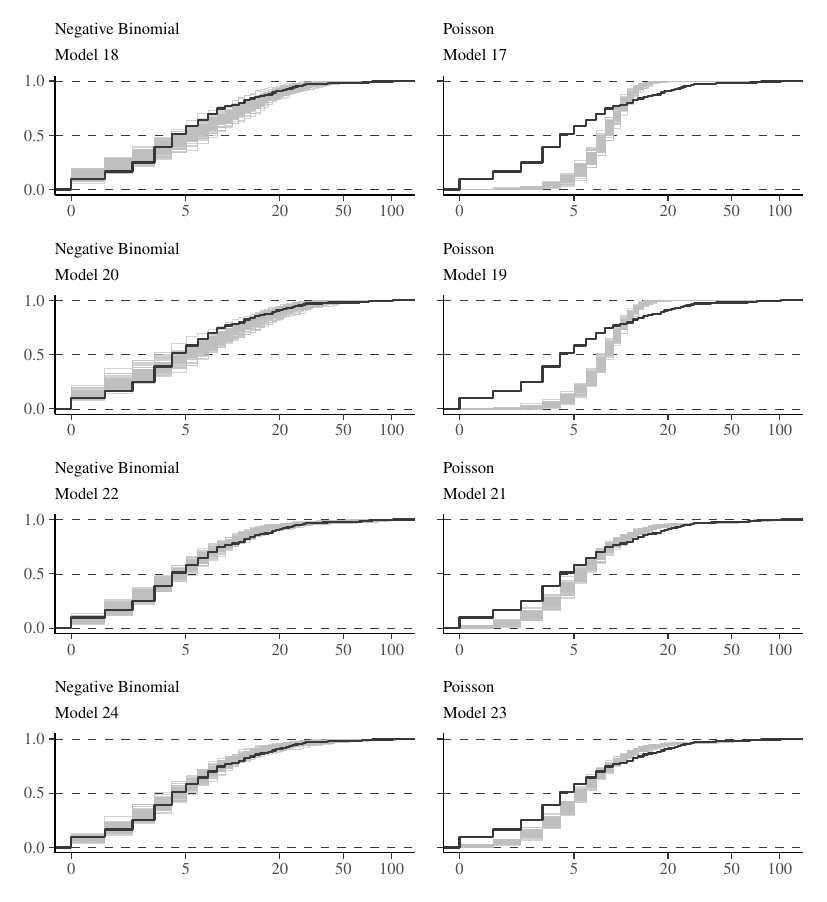}
    \vspace{-0.5cm}
    \caption{Epilepsy case study (Part I). ECDF plots for Model 17 to Model 24. Models in each row only differ with respect to the chosen distributional family for the observations.
    Assuming a Poisson distribution seems to lead to more disagreement with the data compared to the model with negative Binomial distribution.}
    \label{fig:epi-1-initial-ppc-ecdf-model-17-24}
\end{figure}

\clearpage

\subsection{Conditional effects plots for models with interaction effect}\label{appsubsec:cond-effects-casestudy-epi-1-ppc}

\begin{figure}[hbp]
    \centering
    \includegraphics{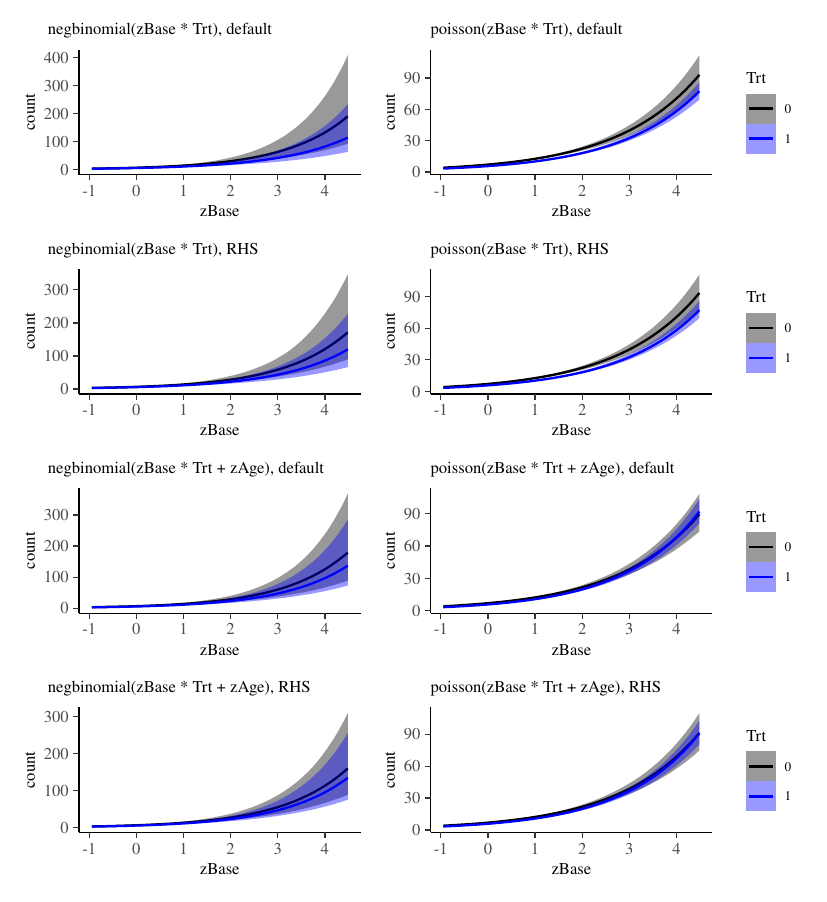}
    \vspace{-0.6cm}
    \caption{Epilepsy case study (Part I). Conditional effects plots for Model 1 to 4 and Model 13 to 16 in Table \ref{tab:models-epi-1-ppc} that include an interaction effect between baseline seizure and treatment.}
    \label{fig:epi-1-initial-cond-effects}
\end{figure}

\clearpage

\section{Epilepsy case study (Part II)}\label{app:casestudy-epi-2-comp}

\subsection{Point-wise differences in estimated \textrm{elpd}}\label{appsubsec:pointwise-elpddiff-casestudy-epi-2-comp}

\begin{figure}[hbp]
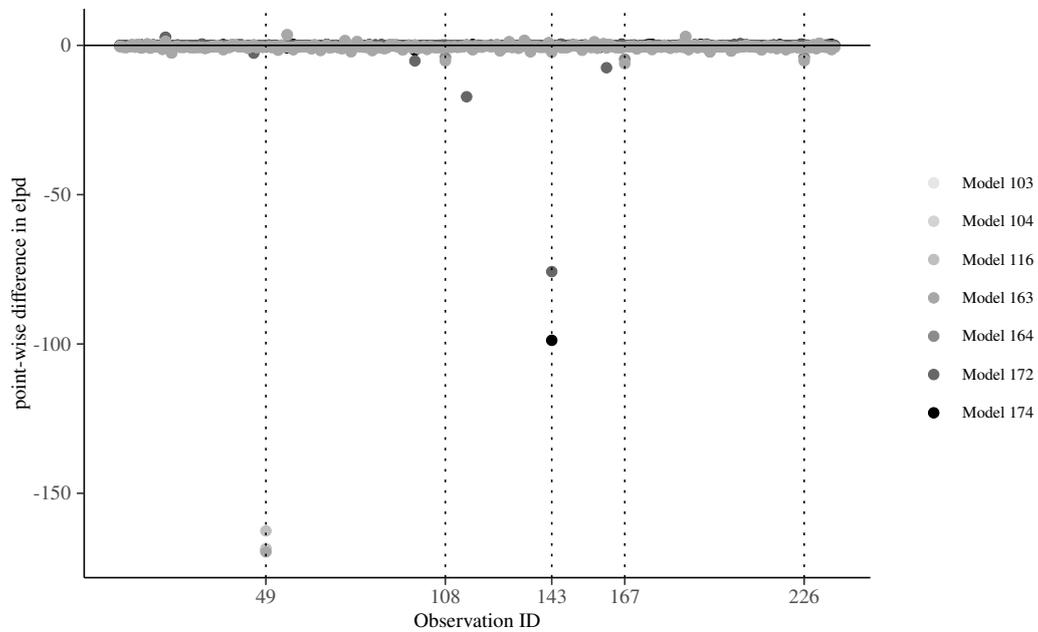

    \centering
    \include{tikz/plot_extended_elpddiff_pointwise_worst_models}
    \caption{Epilepsy case study (Part II). Point-wise differences in \textrm{elpd} estimated with a combination of integrated PSIS-LOO-CV and brute-force LOO-CV for the seven models with lowest $\Delta \widehat{\mathrm{elpd}}$ and large $\widehat{\text{se}}(\Delta \widehat{\mathrm{elpd}})$ in the set of $86$ models. Seizure counts that are $>4 \cdot \sigma$ of the overall observed seizure counts, are indicated with dashed vertical lines and correspond to patient $25$ with observation ID $143$ and patient $49$ with observation IDs $108$, $143$, $167$ and $226$ whose magnitude or pattern of seizure counts differ largely from the rest of the patients.}
    \label{fig:epi-2-pointwise-elpddiff-worst-models}
\end{figure}

\subsection{Modelling choices}\label{appsubsec:models-casestudy-epi-2-comp}
\begin{landscape}
{\small 
\begin{longtable}[c]{llll}
    \caption{Epilepsy case study (Part II). All modelling choices, corresponding Model ID's and model formulae contained in the multiverse of $175$ models with \texttt{brms::brm()} excluding the $17$ models filtered out before. Rows highlighted in \mybox[fill=gray!50]{grey} indicate the filtered set of models as depicted in the right bottom subplot in Figure \ref{fig:epi-2-extended-elpddiff-2se-filtering-combined}.}
    \label{tab:models-epi-2-psisloocv}\\
    \toprule
    Model ID & Obs. family & Priors & Formula\\
    \midrule
    \endfirsthead
    \multicolumn{4}{@{}l}{\textit{(continued)}}\\
    \toprule
    Model ID & Obs. family & Priors & Formula\\
    \midrule
    \endhead

    \endfoot
    \bottomrule
    \endlastfoot
    Model 2 & Neg. Binomial & default in \texttt{brms} & \texttt{count $\sim$ zBase * Trt}\\
    Model 4 & Neg. Binomial & \texttt{brms::horseshoe(3)} & \texttt{count $\sim$ zBase * Trt}\\
    Model 10 & Neg. Binomial & default in \texttt{brms} & \texttt{count $\sim$ Trt+zBase}\\
    Model 12 & Neg. Binomial & \texttt{brms::horseshoe(3)} & \texttt{count $\sim$ Trt+zBase}\\
    Model 16 & Neg. Binomial & \texttt{brms::horseshoe(3)} & \texttt{count $\sim$ zBase * Trt+zAge}\\
    Model 22 & Neg. Binomial & default in \texttt{brms} & \texttt{count $\sim$ Trt+zBase+zAge}\\
    Model 24 & Neg. Binomial & \texttt{brms::horseshoe(3)} & \texttt{count $\sim$ Trt+zBase+zAge}\\
    Model 25 & Poisson & default in \texttt{brms} & \texttt{count $\sim$ zBase * Trt+(1 | patient)}\\
    \rowcolor{lightgray} Model 26 & Neg. Binomial & default in \texttt{brms} & \texttt{count $\sim$ zBase * Trt+(1 | patient)}\\
    \rowcolor{lightgray} Model 27 & Poisson & \texttt{brms::horseshoe(3)} & \texttt{count $\sim$ zBase * Trt+(1 | patient)}\\
    \rowcolor{lightgray} Model 28 & Neg. Binomial & \texttt{brms::horseshoe(3)} & \texttt{count $\sim$ zBase * Trt+(1 | patient)}\\
    \rowcolor{lightgray} Model 29 & Poisson & default in \texttt{brms} & \texttt{count $\sim$ Trt+(1 | patient)}\\
    Model 30 & Neg. Binomial & default in \texttt{brms} & \texttt{count $\sim$ Trt+(1 | patient)}\\
    \rowcolor{lightgray} Model 31 & Poisson & \texttt{brms::horseshoe(3)} & \texttt{count $\sim$ Trt+(1 | patient)}\\
    \rowcolor{lightgray} Model 32 & Neg. Binomial & \texttt{brms::horseshoe(3)} & \texttt{count $\sim$ Trt+(1 | patient)}\\
    Model 33 & Poisson & default in \texttt{brms} & \texttt{count $\sim$ Trt+zBase+(1 | patient)}\\
    Model 34 & Neg. Binomial & default in \texttt{brms} & \texttt{count $\sim$ Trt+zBase+(1 | patient)}\\
    \rowcolor{lightgray} Model 35 & Poisson & \texttt{brms::horseshoe(3)} & \texttt{count $\sim$ Trt+zBase+(1 | patient)}\\
    Model 36 & Neg. Binomial & \texttt{brms::horseshoe(3)} & \texttt{count $\sim$ Trt+zBase+(1 | patient)}\\
    \rowcolor{lightgray} Model 37 & Poisson & default in \texttt{brms} & \texttt{count $\sim$ zBase * Trt+zAge+(1 | patient)}\\
    Model 38 & Neg. Binomial & default in \texttt{brms} & \texttt{count $\sim$ zBase * Trt+zAge+(1 | patient)}\\
    Model 39 & Poisson & \texttt{brms::horseshoe(3)} & \texttt{count $\sim$ zBase * Trt+zAge+(1 | patient)}\\
    Model 40 & Neg. Binomial & \texttt{brms::horseshoe(3)} & \texttt{count $\sim$ zBase * Trt+zAge+(1 | patient)}\\
    \rowcolor{lightgray} Model 41 & Poisson & default in \texttt{brms} & \texttt{count $\sim$ Trt+zAge+(1 | patient)}\\
    Model 42 & Neg. Binomial & default in \texttt{brms} & \texttt{count $\sim$ Trt+zAge+(1 | patient)}\\
    Model 43 & Poisson & \texttt{brms::horseshoe(3)} & \texttt{count $\sim$ Trt+zAge+(1 | patient)}\\
    Model 44 & Neg. Binomial & \texttt{brms::horseshoe(3)} & \texttt{count $\sim$ Trt+zAge+(1 | patient)}\\
    Model 45 & Poisson & default in \texttt{brms} & \texttt{count $\sim$ Trt+zBase+zAge+(1 | patient)}\\
    Model 46 & Neg. Binomial & default in \texttt{brms} & \texttt{count $\sim$ Trt+zBase+zAge+(1 | patient)}\\
    Model 47 & Poisson & \texttt{brms::horseshoe(3)} & \texttt{count $\sim$ Trt+zBase+zAge+(1 | patient)}\\
    Model 48 & Neg. Binomial & \texttt{brms::horseshoe(3)} & \texttt{count $\sim$ Trt+zBase+zAge+(1 | patient)}\\
    Model 49 & Poisson & default in \texttt{brms} & \texttt{count $\sim$ zBase * Trt+(1 | visit)}\\
    Model 50 & Neg. Binomial & default in \texttt{brms} & \texttt{count $\sim$ zBase * Trt+(1 | visit)}\\
    Model 51 & Poisson & \texttt{brms::horseshoe(3)} & \texttt{count $\sim$ zBase * Trt+(1 | visit)}\\
    Model 52 & Neg. Binomial & \texttt{brms::horseshoe(3)} & \texttt{count $\sim$ zBase * Trt+(1 | visit)}\\
    Model 53 & Poisson & default in \texttt{brms} & \texttt{count $\sim$ Trt+(1 | visit)}\\
    Model 54 & Neg. Binomial & default in \texttt{brms} & \texttt{count $\sim$ Trt+(1 | visit)}\\
    Model 55 & Poisson & \texttt{brms::horseshoe(3)} & \texttt{count $\sim$ Trt+(1 | visit)}\\
    Model 56 & Neg. Binomial & \texttt{brms::horseshoe(3)} & \texttt{count $\sim$ Trt+(1 | visit)}\\
    Model 57 & Poisson & default in \texttt{brms} & \texttt{count $\sim$ Trt+zBase+(1 | visit)}\\
    Model 58 & Neg. Binomial & default in \texttt{brms} & \texttt{count $\sim$ Trt+zBase+(1 | visit)}\\
    Model 59 & Poisson & \texttt{brms::horseshoe(3)} & \texttt{count $\sim$ Trt+zBase+(1 | visit)}\\
    Model 60 & Neg. Binomial & \texttt{brms::horseshoe(3)} & \texttt{count $\sim$ Trt+zBase+(1 | visit)}\\
    Model 61 & Poisson & default in \texttt{brms} & \texttt{count $\sim$ zBase * Trt+zAge+(1 | visit)}\\
    Model 62 & Neg. Binomial & default in \texttt{brms} & \texttt{count $\sim$ zBase * Trt+zAge+(1 | visit)}\\
    Model 63 & Poisson & \texttt{brms::horseshoe(3)} & \texttt{count $\sim$ zBase * Trt+zAge+(1 | visit)}\\
    Model 64 & Neg. Binomial & \texttt{brms::horseshoe(3)} & \texttt{count $\sim$ zBase * Trt+zAge+(1 | visit)}\\
    Model 65 & Poisson & default in \texttt{brms} & \texttt{count $\sim$ Trt+zAge+(1 | visit)}\\
    Model 66 & Neg. Binomial & default in \texttt{brms} & \texttt{count $\sim$ Trt+zAge+(1 | visit)}\\
    Model 67 & Poisson & \texttt{brms::horseshoe(3)} & \texttt{count $\sim$ Trt+zAge+(1 | visit)}\\
    Model 68 & Neg. Binomial & \texttt{brms::horseshoe(3)} & \texttt{count $\sim$ Trt+zAge+(1 | visit)}\\
    Model 69 & Poisson & default in \texttt{brms} & \texttt{count $\sim$ Trt+zBase+zAge+(1 | visit)}\\
    Model 70 & Neg. Binomial & default in \texttt{brms} & \texttt{count $\sim$ Trt+zBase+zAge+(1 | visit)}\\
    Model 71 & Poisson & \texttt{brms::horseshoe(3)} & \texttt{count $\sim$ Trt+zBase+zAge+(1 | visit)}\\
    Model 72 & Neg. Binomial & \texttt{brms::horseshoe(3)} & \texttt{count $\sim$ Trt+zBase+zAge+(1 | visit)}\\
    \rowcolor{lightgray} Model 73 & Poisson & default in \texttt{brms} & \texttt{count $\sim$ zBase * Trt+(1 | patient)+(1 | visit)}\\
    Model 74 & Neg. Binomial & default in \texttt{brms} & \texttt{count $\sim$ zBase * Trt+(1 | patient)+(1 | visit)}\\
    \rowcolor{lightgray} Model 75 & Poisson & \texttt{brms::horseshoe(3)} & \texttt{count $\sim$ zBase * Trt+(1 | patient)+(1 | visit)}\\
    Model 76 & Neg. Binomial & \texttt{brms::horseshoe(3)} & \texttt{count $\sim$ zBase * Trt+(1 | patient)+(1 | visit)}\\
    \rowcolor{lightgray} Model 77 & Poisson & default in \texttt{brms} & \texttt{count $\sim$ Trt+(1 | patient)+(1 | visit)}\\
    Model 78 & Neg. Binomial & default in \texttt{brms} & \texttt{count $\sim$ Trt+(1 | patient)+(1 | visit)}\\
    Model 79 & Poisson & \texttt{brms::horseshoe(3)} & \texttt{count $\sim$ Trt+(1 | patient)+(1 | visit)}\\
    Model 80 & Neg. Binomial & \texttt{brms::horseshoe(3)} & \texttt{count $\sim$ Trt+(1 | patient)+(1 | visit)}\\
    \rowcolor{lightgray} Model 81 & Poisson & default in \texttt{brms} & \texttt{count $\sim$ Trt+zBase+(1 | patient)+(1 | visit)}\\
    Model 82 & Neg. Binomial & default in \texttt{brms} & \texttt{count $\sim$ Trt+zBase+(1 | patient)+(1 | visit)}\\
    \rowcolor{lightgray} Model 83 & Poisson & \texttt{brms::horseshoe(3)} & \texttt{count $\sim$ Trt+zBase+(1 | patient)+(1 | visit)}\\
    Model 84 & Neg. Binomial & \texttt{brms::horseshoe(3)} & \texttt{count $\sim$ Trt+zBase+(1 | patient)+(1 | visit)}\\
    \rowcolor{lightgray} Model 85 & Poisson & default in \texttt{brms} & \texttt{count $\sim$ zBase * Trt+zAge+(1 | patient)+(1 | visit)}\\
    Model 86 & Neg. Binomial & default in \texttt{brms} & \texttt{count $\sim$ zBase * Trt+zAge+(1 | patient)+(1 | visit)}\\
    Model 87 & Poisson & \texttt{brms::horseshoe(3)} & \texttt{count $\sim$ zBase * Trt+zAge+(1 | patient)+(1 | visit)}\\
    Model 88 & Neg. Binomial & \texttt{brms::horseshoe(3)} & \texttt{count $\sim$ zBase * Trt+zAge+(1 | patient)+(1 | visit)}\\
    \rowcolor{lightgray} Model 89 & Poisson & default in \texttt{brms} & \texttt{count $\sim$ Trt+zAge+(1 | patient)+(1 | visit)}\\
    Model 90 & Neg. Binomial & default in \texttt{brms} & \texttt{count $\sim$ Trt+zAge+(1 | patient)+(1 | visit)}\\
    \rowcolor{lightgray} Model 91 & Poisson & \texttt{brms::horseshoe(3)} & \texttt{count $\sim$ Trt+zAge+(1 | patient)+(1 | visit)}\\
    Model 92 & Neg. Binomial & \texttt{brms::horseshoe(3)} & \texttt{count $\sim$ Trt+zAge+(1 | patient)+(1 | visit)}\\
    Model 93 & Poisson & default in \texttt{brms} & \texttt{count $\sim$ Trt+zBase+zAge+(1 | patient)+(1 | visit)}\\
    \rowcolor{lightgray}  Model 94 & Neg. Binomial & default in \texttt{brms} & \texttt{count $\sim$ Trt+zBase+zAge+(1 | patient)+(1 | visit)}\\
    \rowcolor{lightgray} Model 95 & Poisson & \texttt{brms::horseshoe(3)} & \texttt{count $\sim$ Trt+zBase+zAge+(1 | patient)+(1 | visit)}\\
    \rowcolor{lightgray} Model 96 & Neg. Binomial & \texttt{brms::horseshoe(3)} & \texttt{count $\sim$ Trt+zBase+zAge+(1 | patient)+(1 | visit)}\\
    Model 97 & Poisson & default in \texttt{brms} & \texttt{count $\sim$ zBase * Trt+(1 | obs)}\\
    Model 98 & Neg. Binomial & default in \texttt{brms} & \texttt{count $\sim$ zBase * Trt+(1 | obs)}\\
    Model 99 & Poisson & \texttt{brms::horseshoe(3)} & \texttt{count $\sim$ zBase * Trt+(1 | obs)}\\
    Model 100 & Neg. Binomial & \texttt{brms::horseshoe(3)} & \texttt{count $\sim$ zBase * Trt+(1 | obs)}\\
    Model 101 & Poisson & default in \texttt{brms} & \texttt{count $\sim$ Trt+(1 | obs)}\\
    Model 102 & Neg. Binomial & default in \texttt{brms} & \texttt{count $\sim$ Trt+(1 | obs)}\\
    Model 103 & Poisson & \texttt{brms::horseshoe(3)} & \texttt{count $\sim$ Trt+(1 | obs)}\\
    Model 104 & Neg. Binomial & \texttt{brms::horseshoe(3)} & \texttt{count $\sim$ Trt+(1 | obs)}\\
    Model 105 & Poisson & default in \texttt{brms} & \texttt{count $\sim$ Trt+zBase+(1 | obs)}\\
    Model 106 & Neg. Binomial & default in \texttt{brms} & \texttt{count $\sim$ Trt+zBase+(1 | obs)}\\
    Model 107 & Poisson & \texttt{brms::horseshoe(3)} & \texttt{count $\sim$ Trt+zBase+(1 | obs)}\\
    Model 108 & Neg. Binomial & \texttt{brms::horseshoe(3)} & \texttt{count $\sim$ Trt+zBase+(1 | obs)}\\
    Model 109 & Poisson & default in \texttt{brms} & \texttt{count $\sim$ zBase * Trt+zAge+(1 | obs)}\\
    Model 110 & Neg. Binomial & default in \texttt{brms} & \texttt{count $\sim$ zBase * Trt+zAge+(1 | obs)}\\
    Model 111 & Poisson & \texttt{brms::horseshoe(3)} & \texttt{count $\sim$ zBase * Trt+zAge+(1 | obs)}\\
    Model 112 & Neg. Binomial & \texttt{brms::horseshoe(3)} & \texttt{count $\sim$ zBase * Trt+zAge+(1 | obs)}\\
    Model 113 & Poisson & default in \texttt{brms} & \texttt{count $\sim$ Trt+zAge+(1 | obs)}\\
    Model 114 & Neg. Binomial & default in \texttt{brms} & \texttt{count $\sim$ Trt+zAge+(1 | obs)}\\
    Model 115 & Poisson & \texttt{brms::horseshoe(3)} & \texttt{count $\sim$ Trt+zAge+(1 | obs)}\\
    Model 116 & Neg. Binomial & \texttt{brms::horseshoe(3)} & \texttt{count $\sim$ Trt+zAge+(1 | obs)}\\
    Model 117 & Poisson & default in \texttt{brms} & \texttt{count $\sim$ Trt+zBase+zAge+(1 | obs)}\\
    Model 118 & Neg. Binomial & default in \texttt{brms} & \texttt{count $\sim$ Trt+zBase+zAge+(1 | obs)}\\
    Model 119 & Poisson & \texttt{brms::horseshoe(3)} & \texttt{count $\sim$ Trt+zBase+zAge+(1 | obs)}\\
    Model 120 & Neg. Binomial & \texttt{brms::horseshoe(3)} & \texttt{count $\sim$ Trt+zBase+zAge+(1 | obs)}\\
    \rowcolor{lightgray} Model 121 & Poisson & default in \texttt{brms} & \texttt{count $\sim$ zBase * Trt+(1 | patient)+(1 | obs)}\\
    Model 122 & Neg. Binomial & default in \texttt{brms} & \texttt{count $\sim$ zBase * Trt+(1 | patient)+(1 | obs)}\\
    Model 123 & Poisson & \texttt{brms::horseshoe(3)} & \texttt{count $\sim$ zBase * Trt+(1 | patient)+(1 | obs)}\\
    Model 124 & Neg. Binomial & \texttt{brms::horseshoe(3)} & \texttt{count $\sim$ zBase * Trt+(1 | patient)+(1 | obs)}\\
    Model 125 & Poisson & default in \texttt{brms} & \texttt{count $\sim$ Trt+(1 | patient)+(1 | obs)}\\
    Model 126 & Neg. Binomial & default in \texttt{brms} & \texttt{count $\sim$ Trt+(1 | patient)+(1 | obs)}\\
    Model 127 & Poisson & \texttt{brms::horseshoe(3)} & \texttt{count $\sim$ Trt+(1 | patient)+(1 | obs)}\\
    Model 128 & Neg. Binomial & \texttt{brms::horseshoe(3)} & \texttt{count $\sim$ Trt+(1 | patient)+(1 | obs)}\\
    Model 129 & Poisson & default in \texttt{brms} & \texttt{count $\sim$ Trt+zBase+(1 | patient)+(1 | obs)}\\
    Model 130 & Neg. Binomial & default in \texttt{brms} & \texttt{count $\sim$ Trt+zBase+(1 | patient)+(1 | obs)}\\
    Model 131 & Poisson & \texttt{brms::horseshoe(3)} & \texttt{count $\sim$ Trt+zBase+(1 | patient)+(1 | obs)}\\
    Model 132 & Neg. Binomial & \texttt{brms::horseshoe(3)} & \texttt{count $\sim$ Trt+zBase+(1 | patient)+(1 | obs)}\\
    Model 133 & Poisson & default in \texttt{brms} & \texttt{count $\sim$ zBase * Trt+zAge+(1 | patient)+(1 | obs)}\\
    Model 134 & Neg. Binomial & default in \texttt{brms} & \texttt{count $\sim$ zBase * Trt+zAge+(1 | patient)+(1 | obs)}\\
    \rowcolor{lightgray} Model 135 & Poisson & \texttt{brms::horseshoe(3)} & \texttt{count $\sim$ zBase * Trt+zAge+(1 | patient)+(1 | obs)}\\
    Model 136 & Neg. Binomial & \texttt{brms::horseshoe(3)} & \texttt{count $\sim$ zBase * Trt+zAge+(1 | patient)+(1 | obs)}\\
    Model 137 & Poisson & default in \texttt{brms} & \texttt{count $\sim$ Trt+zAge+(1 | patient)+(1 | obs)}\\
    Model 138 & Neg. Binomial & default in \texttt{brms} & \texttt{count $\sim$ Trt+zAge+(1 | patient)+(1 | obs)}\\
    \rowcolor{lightgray} Model 139 & Poisson & \texttt{brms::horseshoe(3)} & \texttt{count $\sim$ Trt+zAge+(1 | patient)+(1 | obs)}\\
    Model 140 & Neg. Binomial & \texttt{brms::horseshoe(3)} & \texttt{count $\sim$ Trt+zAge+(1 | patient)+(1 | obs)}\\
    Model 141 & Poisson & default in \texttt{brms} & \texttt{count $\sim$ Trt+zBase+zAge+(1 | patient)+(1 | obs)}\\
    \rowcolor{lightgray} Model 142 & Neg. Binomial & default in \texttt{brms} & \texttt{count $\sim$ Trt+zBase+zAge+(1 | patient)+(1 | obs)}\\
    Model 143 & Poisson & \texttt{brms::horseshoe(3)} & \texttt{count $\sim$ Trt+zBase+zAge+(1 | patient)+(1 | obs)}\\
    Model 144 & Neg. Binomial & \texttt{brms::horseshoe(3)} & \texttt{count $\sim$ Trt+zBase+zAge+(1 | patient)+(1 | obs)}\\
    Model 145 & Poisson & default in \texttt{brms} & \texttt{count $\sim$ zBase * Trt+(1 | visit)+(1 | obs)}\\
    Model 146 & Neg. Binomial & default in \texttt{brms} & \texttt{count $\sim$ zBase * Trt+(1 | visit)+(1 | obs)}\\
    Model 147 & Poisson & \texttt{brms::horseshoe(3)} & \texttt{count $\sim$ zBase * Trt+(1 | visit)+(1 | obs)}\\
    Model 148 & Neg. Binomial & \texttt{brms::horseshoe(3)} & \texttt{count $\sim$ zBase * Trt+(1 | visit)+(1 | obs)}\\
    Model 149 & Poisson & default in \texttt{brms} & \texttt{count $\sim$ Trt+(1 | visit)+(1 | obs)}\\
    Model 150 & Neg. Binomial & default in \texttt{brms} & \texttt{count $\sim$ Trt+(1 | visit)+(1 | obs)}\\
    Model 151 & Poisson & \texttt{brms::horseshoe(3)} & \texttt{count $\sim$ Trt+(1 | visit)+(1 | obs)}\\
    Model 152 & Neg. Binomial & \texttt{brms::horseshoe(3)} & \texttt{count $\sim$ Trt+(1 | visit)+(1 | obs)}\\
    Model 153 & Poisson & default in \texttt{brms} & \texttt{count $\sim$ Trt+zBase+(1 | visit)+(1 | obs)}\\
    Model 154 & Neg. Binomial & default in \texttt{brms} & \texttt{count $\sim$ Trt+zBase+(1 | visit)+(1 | obs)}\\
    Model 155 & Poisson & \texttt{brms::horseshoe(3)} & \texttt{count $\sim$ Trt+zBase+(1 | visit)+(1 | obs)}\\
    Model 156 & Neg. Binomial & \texttt{brms::horseshoe(3)} & \texttt{count $\sim$ Trt+zBase+(1 | visit)+(1 | obs)}\\
    Model 157 & Poisson & default in \texttt{brms} & \texttt{count $\sim$ zBase * Trt+zAge+(1 | visit)+(1 | obs)}\\
    Model 158 & Neg. Binomial & default in \texttt{brms} & \texttt{count $\sim$ zBase * Trt+zAge+(1 | visit)+(1 | obs)}\\
    Model 159 & Poisson & \texttt{brms::horseshoe(3)} & \texttt{count $\sim$ zBase * Trt+zAge+(1 | visit)+(1 | obs)}\\
    Model 160 & Neg. Binomial & \texttt{brms::horseshoe(3)} & \texttt{count $\sim$ zBase * Trt+zAge+(1 | visit)+(1 | obs)}\\
    Model 161 & Poisson & default in \texttt{brms} & \texttt{count $\sim$ Trt+zAge+(1 | visit)+(1 | obs)}\\
    Model 162 & Neg. Binomial & default in \texttt{brms} & \texttt{count $\sim$ Trt+zAge+(1 | visit)+(1 | obs)}\\
    Model 163 & Poisson & \texttt{brms::horseshoe(3)} & \texttt{count $\sim$ Trt+zAge+(1 | visit)+(1 | obs)}\\
    Model 164 & Neg. Binomial & \texttt{brms::horseshoe(3)} & \texttt{count $\sim$ Trt+zAge+(1 | visit)+(1 | obs)}\\
    Model 165 & Poisson & default in \texttt{brms} & \texttt{count $\sim$ Trt+zBase+zAge+(1 | visit)+(1 | obs)}\\
    Model 166 & Neg. Binomial & default in \texttt{brms} & \texttt{count $\sim$ Trt+zBase+zAge+(1 | visit)+(1 | obs)}\\
    Model 167 & Poisson & \texttt{brms::horseshoe(3)} & \texttt{count $\sim$ Trt+zBase+zAge+(1 | visit)+(1 | obs)}\\
    Model 168 & Neg. Binomial & \texttt{brms::horseshoe(3)} & \texttt{count $\sim$ Trt+zBase+zAge+(1 | visit)+(1 | obs)}\\
    \rowcolor{lightgray} Model 169 & Poisson & default in \texttt{brms} & \texttt{count $\sim$ zBase * Trt+(1 | patient)+(1 | visit)+(1 | obs)}\\
    Model 170 & Neg. Binomial & default in \texttt{brms} & \texttt{count $\sim$ zBase * Trt+(1 | patient)+(1 | visit)+(1 | obs)}\\
    Model 171 & Poisson & \texttt{brms::horseshoe(3)} & \texttt{count $\sim$ zBase * Trt+(1 | patient)+(1 | visit)+(1 | obs)}\\
    Model 172 & Neg. Binomial & \texttt{brms::horseshoe(3)} & \texttt{count $\sim$ zBase *   Trt+(1 | patient)+(1 | visit)+(1 | obs)}\\
    Model 173 & Poisson & default in \texttt{brms} & \texttt{count $\sim$ Trt+(1 | patient)+(1 | visit)+(1 | obs)}\\
    Model 174 & Neg. Binomial & default in \texttt{brms} & \texttt{count $\sim$ Trt+(1 | patient)+(1 | visit)+(1 | obs)}\\
    \rowcolor{lightgray} Model 175 & Poisson & \texttt{brms::horseshoe(3)} & \texttt{count $\sim$ Trt+(1 | patient)+(1 | visit)+(1 | obs)}\\
    Model 176 & Neg. Binomial & \texttt{brms::horseshoe(3)} & \texttt{count $\sim$ Trt+(1 | patient)+(1 | visit)+(1 | obs)}\\
    Model 177 & Poisson & default in \texttt{brms} & \texttt{count $\sim$ Trt+zBase+(1 | patient)+(1 | visit)+(1 | obs)}\\
    Model 178 & Neg. Binomial & default in \texttt{brms} & \texttt{count $\sim$ Trt+zBase+(1 | patient)+(1 | visit)+(1 | obs)}\\
    \rowcolor{lightgray} Model 179 & Poisson & \texttt{brms::horseshoe(3)} & \texttt{count $\sim$ Trt+zBase+(1 | patient)+(1 | visit)+(1 | obs)}\\
    Model 180 & Neg. Binomial & \texttt{brms::horseshoe(3)} & \texttt{count $\sim$   Trt+zBase+(1 | patient)+(1 | visit)+(1 | obs)}\\
    \rowcolor{lightgray} Model 181 & Poisson & default in \texttt{brms} & \texttt{count $\sim$ zBase * Trt+zAge+(1 | patient)+(1 | visit)+(1 | obs)}\\
    Model 182 & Neg. Binomial & default in \texttt{brms} & \texttt{count $\sim$ zBase * Trt+zAge+(1 | patient)+(1 | visit)+(1 | obs)}\\
    Model 183 & Poisson & \texttt{brms::horseshoe(3)} & \texttt{count $\sim$ zBase * Trt+zAge+(1 | patient)+(1 | visit)+(1 | obs)}\\
    Model 184 & Neg. Binomial & \texttt{brms::horseshoe(3)} & \texttt{count $\sim$ zBase * Trt+zAge+(1 | patient)+(1 | visit)+(1 | obs)}\\
    Model 185 & Poisson & default in \texttt{brms} & \texttt{count $\sim$ Trt+zAge+(1 | patient)+(1 | visit)+(1 | obs)}\\
    Model 186 & Neg. Binomial & default in \texttt{brms} & \texttt{count $\sim$ Trt+zAge+(1 | patient)+(1 | visit)+(1 | obs)}\\
    Model 187 & Poisson & \texttt{brms::horseshoe(3)} & \texttt{count $\sim$ Trt+zAge+(1 | patient)+(1 | visit)+(1 | obs)}\\
    Model 188 & Neg. Binomial & \texttt{brms::horseshoe(3)} & \texttt{count $\sim$ Trt+zAge+(1 | patient)+(1 | visit)+(1 | obs)}\\
    Model 189 & Poisson & default in \texttt{brms} & \texttt{count $\sim$ Trt+zBase+zAge+(1 | patient)+(1 | visit)+(1 | obs)}\\
    Model 190 & Neg. Binomial & default in \texttt{brms} & \texttt{count $\sim$ Trt+zBase+zAge+(1 | patient)+(1 | visit)+(1 | obs)}\\
    Model 191 & Poisson & \texttt{brms::horseshoe(3)} & \texttt{count $\sim$ Trt+zBase+zAge+(1 | patient)+(1 | visit)+(1 | obs)}\\
    \rowcolor{lightgray} Model 192 & Neg. Binomial & \texttt{brms::horseshoe(3)} & \texttt{count $\sim$ Trt+zBase+zAge+(1 | patient)+(1 | visit)+(1 | obs)}\\*
    \end{longtable}
    }
\end{landscape}

\clearpage

\section{Birthdays (Reparameterisation)}\label{app:casestudy-birthdays}

\subsection{Models}\label{appsubsec:models-casestudy-birthdays}

We use an additive model $y(t) = f_1(t) + f_2(t) + f_3(t) + f_4(t) + f_5(t) + \varepsilon_t$ where

\begin{enumerate}[]
    \item[\textbullet] $t$ is the time in days since 01.01.1969;
    \item[\textbullet] $y_t$ are the registered numbers of births or their natural logarithm depending on the used likelihood;
    \item[\textbullet] $f_1, f_2, f_3, f_4$ and $f_5$ are individual model components and
    \item[\textbullet] $\varepsilon_t$ are the residuals which we assume to be normally distributed.
\end{enumerate}
The multiverse of $144$ models consists of all possible combinations of likelihoods (normal or log-normal) and different additive model components, where the additive model components can be:

\begin{enumerate}[leftmargin=\parindent,align=left]
    \item[$ $] $f_1$ (baseline): 
        \begin{enumerate}[nosep]
            \item[\textbullet] intercept only, $f_1(t) = f_1(0) \sim \mathcal{N}(0,1)$, or
            \item[\textbullet] slowly evolving baseline, $f_1(t) \sim \mathcal{GP}\left(\mathcal{N}(0,1), k_1\right)$ with $$k_1(t, t') = \sigma_1^2\exp\left(-\frac{|t-t'|^2}{2l^2_1}\right)$$.
        \end{enumerate}
    \item[$ $] $f_2$ (seasonal variation):
        \begin{enumerate}[nosep]
            \item[\textbullet] no seasonal variation, $f_2(t) = 0$, or
            \item[\textbullet] yearly periodic GP, $f_2(t) \sim \mathcal{GP}(0, k_2)$ with $$k_2(t, t') = \sigma_2^2\exp\left(-\frac{1}{l^2_2} 2\sin^2(\pi(t-t')/365.25)\right)$$.
        \end{enumerate}
    \item[$ $] $f_3$ (day of the week effect): 
        \begin{enumerate}[nosep]
            \item[\textbullet] no day of the week effect, $f_3(t) = 0$,
            \item[\textbullet] constant day of the week effect, $f_3(t) = f_{3, \mathrm{dow}(t)}$ with $f_{3,i} \sim \mathcal{N}(0,1)$ for $i=1,\cdots,7$, or
            \item[\textbullet] slowly evolving day of the week effect, $f_3(t) = f_{3, \mathrm{dow}(t)} f_{3,GP}(t)$ with $f_{3,GP}(t) \sim \mathcal{GP}(0, k_3)$ with $k_3(t, t') = \sigma_3^2\exp\left(-\frac{|t-t'|^2}{2l^2_3}\right)$
        \end{enumerate}
    \item[$ $] $f_4$ (floating holiday effect):
        \begin{enumerate}[nosep]
            \item[\textbullet] no floating holiday effect, $f_4(t) = 0$, or
            \item[\textbullet] constant floating holiday effect, $f_4(t) = f_{4, \mathrm{floating}(t)}$ with $f_{4, i} \sim \mathcal{N}(0,1)$ overwriting the intercept, the day of the week and the day of the year effect and including the floating holidays Memorial Day, Labour Day and Thanksgiving as separate effects.
        \end{enumerate}
    \item[$ $] $f_5$ (day of the year effect), $(f_5(t) = f_{5,\mathrm{doy}(t)}$ with
        \begin{enumerate}[nosep]
            \item[\textbullet] normal distribution hierarchical prior, $f_{5, i} \sim \mathcal{N}(0, \sigma)$,
            \item[\textbullet] student's t-distribution hierarchical prior, $f_{5, i} \sim t_\nu(0, \sigma)$, or
            \item[\textbullet] regularised horseshoe prior, $f_{5, i} \sim \mathrm{RHS}(0, c, \lambda, \tau)$ with $\nu_c=100$, $\nu_\lambda=1$, $\sigma_\lambda = 2$, $\sigma_\tau = .1$ and $\nu_\tau=100$.
        \end{enumerate}
\end{enumerate}

\subsection{Modelling choices}\label{appsubsec:modelspecs-casestudy-birthdays}

{\small
    \begin{longtable}[c]{lllllll}
    \caption{Birthdays case study (Reparameterisation). All modelling choices resulting in $144$ models with corresponding Model ID's. Rows highlighted in \mybox[fill=gray!50]{grey} indicate the filtered set of models as depicted in Figure \ref{fig:birthdays-elpddiff-combined}.}
    \label{tab:models-birthdays-reparam}\\
    \toprule
    Model ID & Baseline & Seasonal & Day of Week & Floating Holidays & Day of Year & Obs. family \\
    \midrule
    \endfirsthead
    \multicolumn{7}{@{}l}{\textit{(continued)}}\\
    \toprule
    Model ID & Baseline & Seasonal & Day of Week & Floating Holidays & Day of Year & Obs. family \\
    \midrule
    \endhead

    \endfoot
    \bottomrule
    \endlastfoot

Model 1 & constant & none & none & none & normal & normal \\
Model 2 & constant & none & none & none & normal & log normal \\
Model 3 & constant & none & none & none & student's t & normal \\
Model 4 & constant & none & none & none & student's t & log normal \\
Model 5 & constant & none & none & none & RHS & normal \\
Model 6 & constant & none & none & none & RHS & log normal \\
Model 7 & constant & none & none & constant & normal & normal \\
Model 8 & constant & none & none & constant & normal & log normal \\
Model 9 & constant & none & none & constant & student's t & normal \\
Model 10 & constant & none & none & constant & student's t & log normal \\
Model 11 & constant & none & none & constant & RHS & normal \\
Model 12 & constant & none & none & constant & RHS & log normal \\
Model 13 & constant & none & constant & none & normal & normal \\
Model 14 & constant & none & constant & none & normal & log normal \\
Model 15 & constant & none & constant & none & student's t & normal \\
Model 16 & constant & none & constant & none & student's t & log normal \\
Model 17 & constant & none & constant & none & RHS & normal \\
Model 18 & constant & none & constant & none & RHS & log normal \\
Model 19 & constant & none & constant & constant & normal & normal \\
Model 20 & constant & none & constant & constant & normal & log normal \\
Model 21 & constant & none & constant & constant & student's t & normal \\
Model 22 & constant & none & constant & constant & student's t & log normal \\
Model 23 & constant & none & constant & constant & RHS & normal \\
Model 24 & constant & none & constant & constant & RHS & log normal \\
Model 25 & constant & none & evolving & none & normal & normal \\
Model 26 & constant & none & evolving & none & normal & log normal \\
Model 27 & constant & none & evolving & none & student's t & normal \\
Model 28 & constant & none & evolving & none & student's t & log normal \\
Model 29 & constant & none & evolving & none & RHS & normal \\
Model 30 & constant & none & evolving & none & RHS & log normal \\
Model 31 & constant & none & evolving & constant & normal & normal \\
Model 32 & constant & none & evolving & constant & normal & log normal \\
Model 33 & constant & none & evolving & constant & student's t & normal \\
Model 34 & constant & none & evolving & constant & student's t & log normal \\
Model 35 & constant & none & evolving & constant & RHS & normal \\
Model 36 & constant & none & evolving & constant & RHS & log normal \\
Model 37 & constant & periodic & none & none & normal & normal \\
Model 38 & constant & periodic & none & none & normal & log normal \\
Model 39 & constant & periodic & none & none & student's t & normal \\
Model 40 & constant & periodic & none & none & student's t & log normal \\
Model 41 & constant & periodic & none & none & RHS & normal \\
Model 42 & constant & periodic & none & none & RHS & log normal \\
Model 43 & constant & periodic & none & constant & normal & normal \\
Model 44 & constant & periodic & none & constant & normal & log normal \\
Model 45 & constant & periodic & none & constant & student's t & normal \\
Model 46 & constant & periodic & none & constant & student's t & log normal \\
Model 47 & constant & periodic & none & constant & RHS & normal \\
Model 48 & constant & periodic & none & constant & RHS & log normal \\
Model 49 & constant & periodic & constant & none & normal & normal \\
Model 50 & constant & periodic & constant & none & normal & log normal \\
Model 51 & constant & periodic & constant & none & student's t & normal \\
Model 52 & constant & periodic & constant & none & student's t & log normal \\
Model 53 & constant & periodic & constant & none & RHS & normal \\
Model 54 & constant & periodic & constant & none & RHS & log normal \\
Model 55 & constant & periodic & constant & constant & normal & normal \\
Model 56 & constant & periodic & constant & constant & normal & log normal \\
Model 57 & constant & periodic & constant & constant & student's t & normal \\
Model 58 & constant & periodic & constant & constant & student's t & log normal \\
Model 59 & constant & periodic & constant & constant & RHS & normal \\
Model 60 & constant & periodic & constant & constant & RHS & log normal \\
Model 61 & constant & periodic & evolving & none & normal & normal \\
Model 62 & constant & periodic & evolving & none & normal & log normal \\
Model 63 & constant & periodic & evolving & none & student's t & normal \\
Model 64 & constant & periodic & evolving & none & student's t & log normal \\
Model 65 & constant & periodic & evolving & none & RHS & normal \\
Model 66 & constant & periodic & evolving & none & RHS & log normal \\
Model 67 & constant & periodic & evolving & constant & normal & normal \\
Model 68 & constant & periodic & evolving & constant & normal & log normal \\
Model 69 & constant & periodic & evolving & constant & student's t & normal \\
Model 70 & constant & periodic & evolving & constant & student's t & log normal \\
Model 71 & constant & periodic & evolving & constant & RHS & normal \\
Model 72 & constant & periodic & evolving & constant & RHS & log normal \\
Model 73 & evolving & none & none & none & normal & normal \\
Model 74 & evolving & none & none & none & normal & log normal \\
Model 75 & evolving & none & none & none & student's t & normal \\
Model 76 & evolving & none & none & none & student's t & log normal \\
Model 77 & evolving & none & none & none & RHS & normal \\
Model 78 & evolving & none & none & none & RHS & log normal \\
Model 79 & evolving & none & none & constant & normal & normal \\
Model 80 & evolving & none & none & constant & normal & log normal \\
Model 81 & evolving & none & none & constant & student's t & normal \\
Model 82 & evolving & none & none & constant & student's t & log normal \\
Model 83 & evolving & none & none & constant & RHS & normal \\
Model 84 & evolving & none & none & constant & RHS & log normal \\
Model 85 & evolving & none & constant & none & normal & normal \\
Model 86 & evolving & none & constant & none & normal & log normal \\
Model 87 & evolving & none & constant & none & student's t & normal \\
Model 88 & evolving & none & constant & none & student's t & log normal \\
Model 89 & evolving & none & constant & none & RHS & normal \\
Model 90 & evolving & none & constant & none & RHS & log normal \\
Model 91 & evolving & none & constant & constant & normal & normal \\
Model 92 & evolving & none & constant & constant & normal & log normal \\
Model 93 & evolving & none & constant & constant & student's t & normal \\
Model 94 & evolving & none & constant & constant & student's t & log normal \\
Model 95 & evolving & none & constant & constant & RHS & normal \\
Model 96 & evolving & none & constant & constant & RHS & log normal \\
Model 97 & evolving & none & evolving & none & normal & normal \\
Model 98 & evolving & none & evolving & none & normal & log normal \\
Model 99 & evolving & none & evolving & none & student's t & normal \\
Model 100 & evolving & none & evolving & none & student's t & log normal \\
Model 101 & evolving & none & evolving & none & RHS & normal \\
Model 102 & evolving & none & evolving & none & RHS & log normal \\
Model 103 & evolving & none & evolving & constant & normal & normal \\
Model 104 & evolving & none & evolving & constant & normal & log normal \\
Model 105 & evolving & none & evolving & constant & student's t & normal \\
Model 106 & evolving & none & evolving & constant & student's t & log normal \\
Model 107 & evolving & none & evolving & constant & RHS & normal \\
Model 108 & evolving & none & evolving & constant & RHS & log normal \\
Model 109 & evolving & periodic & none & none & normal & normal \\
Model 110 & evolving & periodic & none & none & normal & log normal \\
Model 111 & evolving & periodic & none & none & student's t & normal \\
Model 112 & evolving & periodic & none & none & student's t & log normal \\
Model 113 & evolving & periodic & none & none & RHS & normal \\
Model 114 & evolving & periodic & none & none & RHS & log normal \\
Model 115 & evolving & periodic & none & constant & normal & normal \\
Model 116 & evolving & periodic & none & constant & normal & log normal \\
Model 117 & evolving & periodic & none & constant & student's t & normal \\
Model 118 & evolving & periodic & none & constant & student's t & log normal \\
Model 119 & evolving & periodic & none & constant & RHS & normal \\
Model 120 & evolving & periodic & none & constant & RHS & log normal \\
Model 121 & evolving & periodic & constant & none & normal & normal \\
Model 122 & evolving & periodic & constant & none & normal & log normal \\
Model 123 & evolving & periodic & constant & none & student's t & normal \\
Model 124 & evolving & periodic & constant & none & student's t & log normal \\
Model 125 & evolving & periodic & constant & none & RHS & normal \\
Model 126 & evolving & periodic & constant & none & RHS & log normal \\
Model 127 & evolving & periodic & constant & constant & normal & normal \\
Model 128 & evolving & periodic & constant & constant & normal & log normal \\
Model 129 & evolving & periodic & constant & constant & student's t & normal \\
Model 130 & evolving & periodic & constant & constant & student's t & log normal \\
Model 131 & evolving & periodic & constant & constant & RHS & normal \\
Model 132 & evolving & periodic & constant & constant & RHS & log normal \\
Model 133 & evolving & periodic & evolving & none & normal & normal \\
Model 134 & evolving & periodic & evolving & none & normal & log normal \\
Model 135 & evolving & periodic & evolving & none & student's t & normal \\
Model 136 & evolving & periodic & evolving & none & student's t & log normal \\
Model 137 & evolving & periodic & evolving & none & RHS & normal \\
Model 138 & evolving & periodic & evolving & none & RHS & log normal \\
Model 139 & evolving & periodic & evolving & constant & normal & normal \\
Model 140 & evolving & periodic & evolving & constant & normal & log normal \\
Model 141 & evolving & periodic & evolving & constant & student's t & normal \\
\rowcolor{lightgray} Model 142 & evolving & periodic & evolving & constant & student's t & log normal \\
Model 143 & evolving & periodic & evolving & constant & RHS & normal \\
\rowcolor{lightgray} Model 144 & evolving & periodic & evolving & constant & RHS & log normal \\
    
    \end{longtable}
}

\end{appendices}
   
\end{document}